\documentclass[useAMS,usenatbib]{mn2e}

\usepackage{times}
\usepackage{color}

\usepackage{amsmath}
\usepackage{textcomp}
\usepackage{amssymb}
\usepackage{graphicx}
\graphicspath{{images/}}
\usepackage{subfigure} 
\usepackage[T1]{fontenc} 
 \usepackage{aecompl}

\title[Are pulsars born with a hidden magnetic field?]
{Are pulsars born with a hidden magnetic field?}
\author[A. Torres-Forn\'e, P. Cerd\'a-Dur\'an, J.A. Pons and J. A. Font]
{Alejandro Torres-Forn\'e$^{1}$\thanks{E-mail: Alejandro.Torres@uv.es}, 
 Pablo Cerd\'a-Dur\'an$^1$, Jos\'e A. Pons$^2$ and Jos\'e A. Font$^{1,3}$\\
$^{1}$Departamento de Astronom\'ia y Astrof\'isica, Universitat de Val\`encia, Dr. Moliner 50, 46100, Burjassot (Val\`encia), Spain\\
$^{2}$Departament de F\'isica Aplicada, Universitat d'Alacant, Ap. Correus 99, 03080 Alacant, Spain\\
$^{3}$Observatori Astron\`omic, Universitat de Val\`encia, Catedr\'atico Jos\'e Beltr\'an 2, 46980, Paterna (Val\`encia), Spain}

\begin{document}


\maketitle

\label{firstpage}

\begin{abstract}
The observation of several neutron stars in the center of supernova remnants and with significantly lower values of the dipolar magnetic field than the average radio-pulsar population has motivated a lively debate about their formation and origin, with controversial interpretations. A possible explanation requires the slow rotation of the proto-neutron star at birth, which is unable to amplify its magnetic field to typical pulsar levels. An alternative possibility, the hidden magnetic field scenario, considers the accretion of the fallback of the supernova debris onto the neutron star as responsible for the submergence (or screening) of the field and its apparently low value. In this paper we study under which conditions the magnetic field of a neutron star can be buried into the crust due to an accreting, conducting fluid. For this purpose, we consider a spherically symmetric  calculation in general relativity to estimate the balance between the incoming accretion flow and the magnetosphere. Our study analyses several models with different specific entropy, composition, and neutron star masses.
The main conclusion of our work is that typical magnetic fields of a few times $10^{12}$ G can be buried by accreting only $10^{-3}-10^{-2} {M}_\odot$,  a relatively modest amount of mass. In view of this result,  the Central Compact Object scenario should not be considered unusual, and we predict that anomalously weak magnetic fields should be common in very young ($<$ few kyr) neutron stars.
\end{abstract}

\begin{keywords}
stars: magnetic field -- stars: neutron -- pulsars: general
\end{keywords}
\section{Introduction}

Central Compact Objects (CCOs) are isolated, young neutron stars (NSs) which  show no radio emission and are located near the center of young supernova remnants (SNRs). Three such NSs, PSR E1207.4-5209, PSR J0821.0-4300, and PSR J1852.3-0040, show an inferred magnetic field significantly lower than the standard values for radio-pulsars (i.e.~$10^{12}$ G). The main properties of these sources are summarized in Table~\ref{Table1}. In all cases, the difference between the characteristic age of the neutron star 
$\tau_c=P/\dot{P}$ and the age of the SNR indicates that these NSs were born spinning at nearly their present periods ($P\sim 0.1-0.4$~s). This discovery has challenged theoretical models of magnetic field generation, that need to be modified to account for their peculiar properties.

\begin{table*}
 \centering
 \begin{minipage}{140mm}
  \caption{Central Compact Objects in Supernova Remnants. From left to right the columns indicate the name of the CCO, the age, the distance $d$, the period $P$, the inferred surface magnetic field, $\rm{B}_{\rm{s}}$, the bolometric luminosity in X-rays,  $\rm{L}_{x,bol}$, the name of the remnant, the characteristic age, and bibliographical references.}
  \begin{tabular}{@{}lcccccccc@{}}
  \hline
   \multicolumn{1}{c}{CCO}    &        Age &d& P& $\rm{B}_{\rm{s}}$
     & $\rm{L}_{x,bol}$ & SNR    & $\tau_c$ &References  \\
        &  (kyr)  & (kpc)  & (s) & $10^{11} $G & (erg $\rm{s}^{-1})$& &(Myr)&\\
        \hline
       		J0822.0-4300& 3.7 &2.2 &0.112 &$0.65$&$6.5\times10 ^{33}$&Puppis A &$190$ &1, 2 \\
		1E 1207.4-5209 & 7 &2.2& 0.424&$2$&$2.5\times10^{33}$&PKS 1209-51/52 &$ 310$ &2, 3, 4, 5, 6, 7\\
		 J185238.6 + 004020 & 7& 7& 0.105& 0.61 &$5.3\times10^{33}$&Kes 79&$ 190$&8, 9, 10, 11\\
\hline
\label{Table1}
\end{tabular}
{\bf References:} (1) \cite{Hui:2006}, (2) \cite{Gotthelf:2013}, (3) \cite{Zavlin:2000}, (4) \cite{Mereghetti:2002}, (5) \cite{Bignami:2003}, (6) \cite{DeLuca:2004}, (7) \cite{Gottfeld:2007}, (8) \cite{Seward:2003}, (9) \cite{Gottfeld:2005}, (10) \cite{Halpern:2007}, (11) \cite{Halpern:2010} 
\end{minipage}
\end{table*}

The first possible explanation for the unusual magnetic field found in these objects simply assumes that these NSs are born with 
a magnetic field much lower than that of their classmates. This value can be amplified by
turbulent dynamo action during the proto-neutron star (PNS) phase \citep{Thompson:1993,Bonanno:2005} . In this model, the final low values of the magnetic field would reflect the fact that the slow rotation of the neutron star at birth does not suffice to effectively amplify the magnetic field through dynamo effects. However, recent studies have shown that, even in the absence of rapid rotation, magnetic fields in PNS can be amplified by other mechanisms such as convection and the standing accretion shock instability (SASI) \citep{Endeve:2012, Obergaulinger:2014}. 

An alternative explanation is the hidden magnetic field scenario \citep{Young:1995, Muslimov:1995, Geppert:1999,Shabaltas:2012}. Following the supernova explosion and the neutron star birth, the supernova shock travels outwards through the external layers of the star. When this shock crosses a discontinuity in density, it is partially reflected and moves backwards (reverse shock). The total mass accreted by the reverse shock in this process is in the range from $\sim 10^{-4} M_\odot$ to a few solar masses on a typical timescale of hours to days \citep{Ugliano:2012}.  Such a high accretion rate can compress the magnetic field of the NS which can eventually be buried into the neutron star crust. As a result, the value of the external magnetic field would be significantly lower than the internal `hidden' magnetic field. \cite{Bernal:2010} performed 1D and 2D numerical simulations of a single column of material falling onto a magnetized neutron star and showed how the magnetic field can be buried into the neutron star crust.

Once the accretion process stops, the magnetic field might eventually reemerge.
The initial studies investigated the process of reemergence using simplified 1D models and dipolar fields \citep{Young:1995, Muslimov:1995, Geppert:1999} and established that the timescale for the magnetic field reemergence is $\sim1-10^{7}$~kyr, 
critically depending on the depth at which the magnetic field is buried.  
More recent investigations have confirmed this result. \cite{Ho:2011} observed similar timescales for the reemergence using a 1D cooling code.  \cite{Vigano:2012} carried out simulations of the evolution of the interior magnetic field during the accretion phase and the  magnetic field submergence phase. 

In the present work we study the feasibility of the hidden magnetic field scenario using a novel numerical approach based 
on the solutions of 1D Riemann problems (discontinuous initial value problems) to model the compression of the magnetic field of the NS.
The two initial states for the Riemann problem are defined by the magnetosphere and by the accreting fluid, at either sides of a moving, discontinuous interface. Following the notation defined in \cite{Michel:1977}, the NS magnetosphere refers to the area surrounding the star where the magnetic pressure dominates over the thermal pressure of the accreting fluid. The magnetopause is the interface between the magnetically dominated area and the thermally dominated area. The equilibrium point is defined as the radius at which the velocity of the contact discontinuity is zero. 

The paper is organized as follows. In Sections \ref{Model} to \ref{sec:magnetosphere} we present the model we use to perform our study. We describe in these sections the equation of state (EoS) of the accreting fluid, the spherically symmetric Michel solution characterizing the accreting fluid, and all the expressions needed to compute the potential solution for the magnetic field in the magnetosphere. Section \ref{Results} contains the main results  of this work. After establishing a reference model, we vary the remaining parameters, namely entropy, composition and the initial distribution of the magnetic field, and study their influence on the fate of the magnetic field. Finally, in Section \ref{Summary} we summarize the main results of our study and present our conclusions and plans for future work.
If not explicitly stated otherwise we use units of $G=c=1$. Greek indices ($\mu,\nu \dots$) run from $0$ to $3$ and latin indices ($i, j \dots$)
form $1$ to $3$.

\section{The reverse shock and the fallback scenario.}
\label{Model}

At the end of their lives, massive stars ($M_{\rm star} \gtrsim 8 M_{\odot}$) possess an onion-shell structure as a result of successive stages of nuclear burning. An inner core, typically formed by iron, with a mass of $\sim 1.4M_{\odot}$ and $\sim 1000$~km radius develops at the centre, balancing gravity through the pressure generated by a relativistic, degenerate, $\gamma=4/3$, fermion gas.
The iron core is unstable due to photo-disintegration of nuclei and electron captures, which result in a deleptonization of the core and a significant pressure reduction ($\gamma<4/3$). As a result, the core shrinks and collapses gravitationally to nuclear matter densities on dynamical timescales ($\sim 100$~ms). As the center of the star reaches nuclear saturation density ($\sim 2\times 10^{14}$~g~cm$^{-3}$), the EoS stiffens and an outward moving (prompt) shock is produced.
As it propagates out the shock suffers severe energy losses dissociating Fe nuclei into free
nucleons ($\sim 1.7 \times 10^{51}$~erg$/0.1 M_{\odot}$), consuming its entire kinetic energy inside the iron core (it stalls at $\sim 100-200$~km), becoming a standing accretion shock in a few ms. There is still debate about the exact mechanism and conditions
for a successful explosion, but it is commonly accepted that the standing shock has to be revived on a timescale of $\lesssim 1$~s by the energy deposition of neutrinos streaming out of the innermost regions, and some form of convective transport for
the shock to carry sufficient energy to disrupt the whole star \citep[see][for a review on the topic]{Janka:2007}. 

Even if the shock is sufficiently strong to power the supernova, part of the material between the nascent neutron star and the propagating shock may fall back into the neutron star \citep{Colgate:1971,Chevalier:1989}. Determining the amount of fallback material depends not only on the energy of the shock but also on the radial structure of the progenitor star \citep{Fryer:2006}. Most of the fallback accretion is the result of the formation of an inward moving reverse shock produced as the main supernova-driving shock crosses the discontinuity between the helium shell and the hydrogen envelope
\citep{Chevalier:1989}. For typical supernova progenitors ($10-30\,M_\odot$) the base of the
hydrogen envelope is at $r_{\rm H} \sim 10^{11}$~cm to $3\times 10^{12}$~cm \citep{Woosley:2002}, which is reached by the main shock on a timescale of a few hours. The reverse shock travels inwards carrying mass that accretes onto the NS. It reaches the vicinity of the NS on a timescale of hours, about the same time at which the main supernova shock reaches the surface of the star \citep{Chevalier:1989}. 
By the time the reverse shock reaches the NS, the initially hot proto-neutron star has cooled down significantly. In its first minute of life the PNS contracts, cools down to $T < 10^{10}$~K and becomes transparent to neutrinos \citep{BL:1986, Pons:1999}. In the next few hours the inner crust ($\rho \in [2\times 10^{11},2\times 10^{14}]$~g~cm$^{-3}$) solidifies but the low density envelope ($\rho < 2\times10^{11}$~g~cm$^{-3}$), which will form the outer crust on a timescale of $1-100$~yr, remains fluid \citep{Page:2004,Aguilera:2008}. 

Understanding the processes generating the magnetic field observed in NSs, in the range from $\sim 10^{10}$~G to $\sim 10^{15}$~G, is still a open issue. Most likely, convection, rotation and turbulence during the PNS phase play a crucial role in field amplification \citep{Thompson:1993}. However, at the time in the evolution that we are considering (hours after birth), none of these processes can be active anymore and the electric current distribution generating the magnetic field will be frozen in the interior of the NS. These currents evolve now on the characteristic Hall and Ohmic timescales of $10^4$-$10^6$~yr \citep{Pons:2007, Pons:2009, Vigano:2013}, much longer than the timescale $t_{\rm acc}$ during which fallback is significant, which 
can be estimated as the free-fall time from the base of the hydrogen envelope
\begin{eqnarray}
t_{\rm acc} \sim 
 \frac{1}{2}\left(\frac{r_{\rm H}^3}{GM}\right)^{1/2}\,.
\end{eqnarray}
This ranges from 30 minutes to several days for the typical values of $r_{\rm H}$ and a $M=1.4 M_{\odot}$.

The total mass accreted during this phase is more uncertain. Detailed 1D numerical simulations of the shock propagation and fallback estimate that typical values range from $10^{-4} M_\odot$ to a few solar masses \citep{Woosley:1995,Zhang:2008,Ugliano:2012}. If more than a solar mass is accreted, the final outcome would be the delayed
formation of a black hole, hours to days after core bounce. 
\cite{Chevalier:1989} and \cite{Zhang:2008} showed that the accretion rate is expected to be maximum when the reverse shock reaches the NS and decreases as $t^{-5/3}$ at later times. Therefore, the total amount of accreted mass is dominated by the fallback during the first few hours. Given the theoretical uncertainties, we assume for the rest of this work that a total mass of $\delta M \in [10^{-5}  M_\odot,\delta M_{\rm max}]$ is accreted during a typical timescale of $t_{\rm acc} \in [10^3,10^4]$~s, being $\delta M_{\rm max}\sim 1 M_\odot$ the amount of mass necessary to add to the NS to form a black hole.
Therefore, the typical accretion rate during fallback is $\dot M \in [10^{-9}, 10^{-3}] M_\odot / \rm{s}$, which, for practical purposes, we assume to stay constant during the accretion phase. This accretion rate, even at its lowest value, exceeds by far the Eddington luminosity 
\begin{eqnarray}
  \frac{{\dot M} c^2}{L_{\rm Edd}} = 5\times 10^6 \left ( \frac{\dot M}{10^{-9} M_\odot /	\rm{s}} \right ),
\end{eqnarray}
with $L_{\rm Edd} = 3.5\times 10^{38}$~erg~s$^{-1}$ the Eddington luminosity for electron scattering. 

In the hypercritical accretion regime, the optical depth is so large
that photons are advected inwards with the flow faster than they can diffuse outwards \citep{Blondin:1986, Chevalier:1989, Houck:1991}. As a result the accreting material cannot cool down resulting in an adiabatic compression of the  fluid. The dominant process cooling down the accreting fluid and releasing the energy stored in the infalling fluid is neutrino emission \citep{Houck:1991}. At temperatures above the pair creation threshold, $T_{\rm pair}\approx 10^{10}$~K, pair annihilation can produce neutrino-antineutrino pairs, for which the infalling material is essentially transparent and are able to cool down very efficiently the material as it is decelerated at the surface of the NS or at the magnetopause. Therefore, the specific entropy, $s$, of the fallback material remains constant all through the accretion phase until it decelerates in the vicinity of the NS.

The value for $s$ is set at the time of the reverse shock formation. Detailed 2D numerical simulations of the propagation of the shock through the star \citep{Scheck:2006,Kifonidis:2003,Kifonidis:2006} show that typical values of $s\sim20~ k_{\rm B}/{\rm nuc}$ are found at the
reverse shock. At this stage of the explosion the flow is highly anisotropic due to the Rayleigh-Taylor instability present in the expanding material and the Richtmyer-Meshkov instability at the He/H interface. Those instabilities generate substantial mixing between hydrogen and helium
and even clumps of high-entropy heavier elements (from C to Ni) rising from the innermost parts of the star. Therefore, the fallback material has entropy in the range $s \sim 1-100 ~k_{\rm B}/{\rm nuc}$ and its composition, although it is mostly helium, can contain almost any element present in the explosion. 3D simulations show qualitatively similar results regarding the entropy values and mixing \citep{Hammer:2010, Joggerst:2010, Wongwathanarat:2015}. 

Outside the NS, the expanding supernova explosion leaves behind a low density rarefaction wave which is rapidly filled by the NS magnetic field, 
forming the  magnetosphere. For the small magnetospheric densities, the inertia of the fluid can be neglected, and the magnetosphere 
can be considered force-free. The fallback reverse shock propagates inwards compressing this magnetosphere. The boundary between the unmagnetized material falling back and the force-free magnetosphere, i.e. the magnetopause, can be easily compressed at long distances ($r \gtrsim 10^8$~cm ) due to the large difference of the pressure of the infalling material with respect to the magnetic pressure. The dynamical effect of the magnetosphere only plays a role at
$r \lesssim 10^8$~cm, i.e. inside the light cylinder for most cases. The precise radius where the magnetic field becomes dynamically relevant is estimated later in Section \ref{sec:riemann}. Only in the case of magnetar-like magnetic fields  and fast initial spin ($P\lesssim 10$~ms) this consideration is not valid, although this is not the case
for CCOs.

To conclude this scenario overview, we note that the magnetospheric torques will spin-down the NS on 
a characteristic timescale \citep{Shapiro:2004} given by 
\begin{eqnarray}
  \tau_c = \frac{P}{2 \dot P} \sim 180 \left( \frac{B_p}{10^{15} {\rm G}} \right)^{-2} \left ( \frac{P}{1 {\rm s}}\right )^2 {\rm yr},
\end{eqnarray}
for a typical NS with radius $10$~km and mass $1.4 M_\odot$. $B_p$ is the value of the magnetic field at the pole of the NS. The value of the moment of inertia is $1.4\times10^{45}~\rm{g}~\rm{cm}^2$.
At birth, the spin period of a NS is limited by the mass-shedding limit to be $P>1$~ms  \citep{Goussard:1998}. 
If all NSs were born with millisecond periods, purely magneto-dipolar spin-down would limit the observed period
of young NSs ($10^4$~yr) to
\begin{eqnarray}
  P_{{\rm obs}, 10^4 {\rm yr}}\lesssim 5.5 \left ( \frac{B}{10^{15} {\rm G}}\right )^2 {\rm s}.
\end{eqnarray}
For magnetic fields $B\lesssim 1.4\times 10^{13}$~G this criterion fails for the vast majority of
pulsars and all CCOs ($P\gtrsim0.1$~s) and therefore the measured spin period must be now very close to that hours after the onset of the supernova explosion. Detailed population synthesis studies of the radio-pulsar population clearly favor a broad initial period distribution in the range 0.1-0.5 s 
\citep{Faucher:2006, Gullon:2014}, rather than fast millisecond pulsars. Therefore, from observational constraints, it is reasonable 
to assume that progenitors of pulsars (including CCOs) have spin periods of $P\sim 0.1-0.5$~s at the moment of fallback.
For such low rotation rates, the NS can be safely considered as a spherically symmetric body and its structure can thus be computed by solving the Tolman-Oppenheimer-Volkoff (TOV) equation.

\begin{figure*}
	\begin{subfigure}
	{\includegraphics[height=55mm]{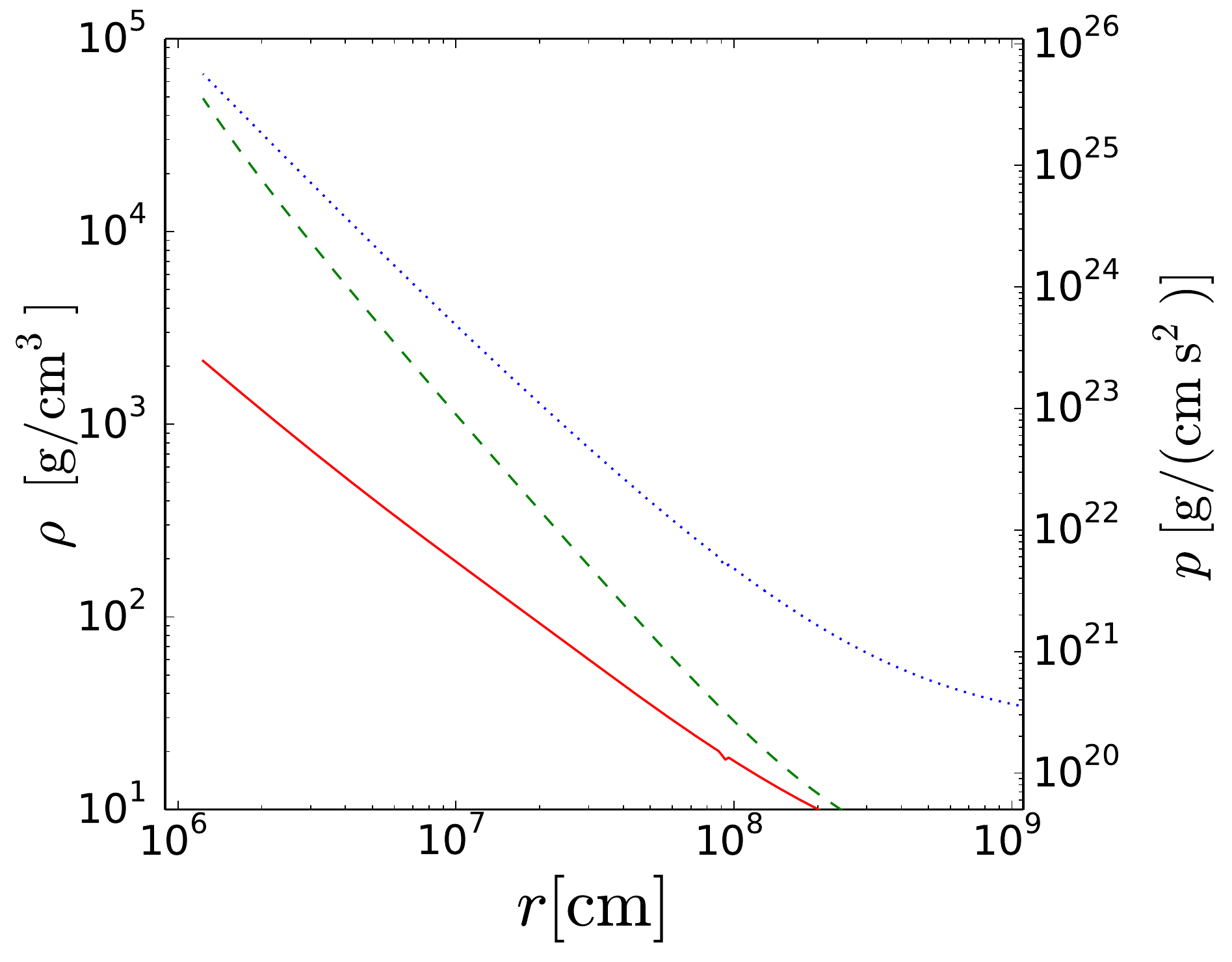}}
	\end{subfigure}
	\begin{subfigure}
	{\includegraphics[height=55mm]{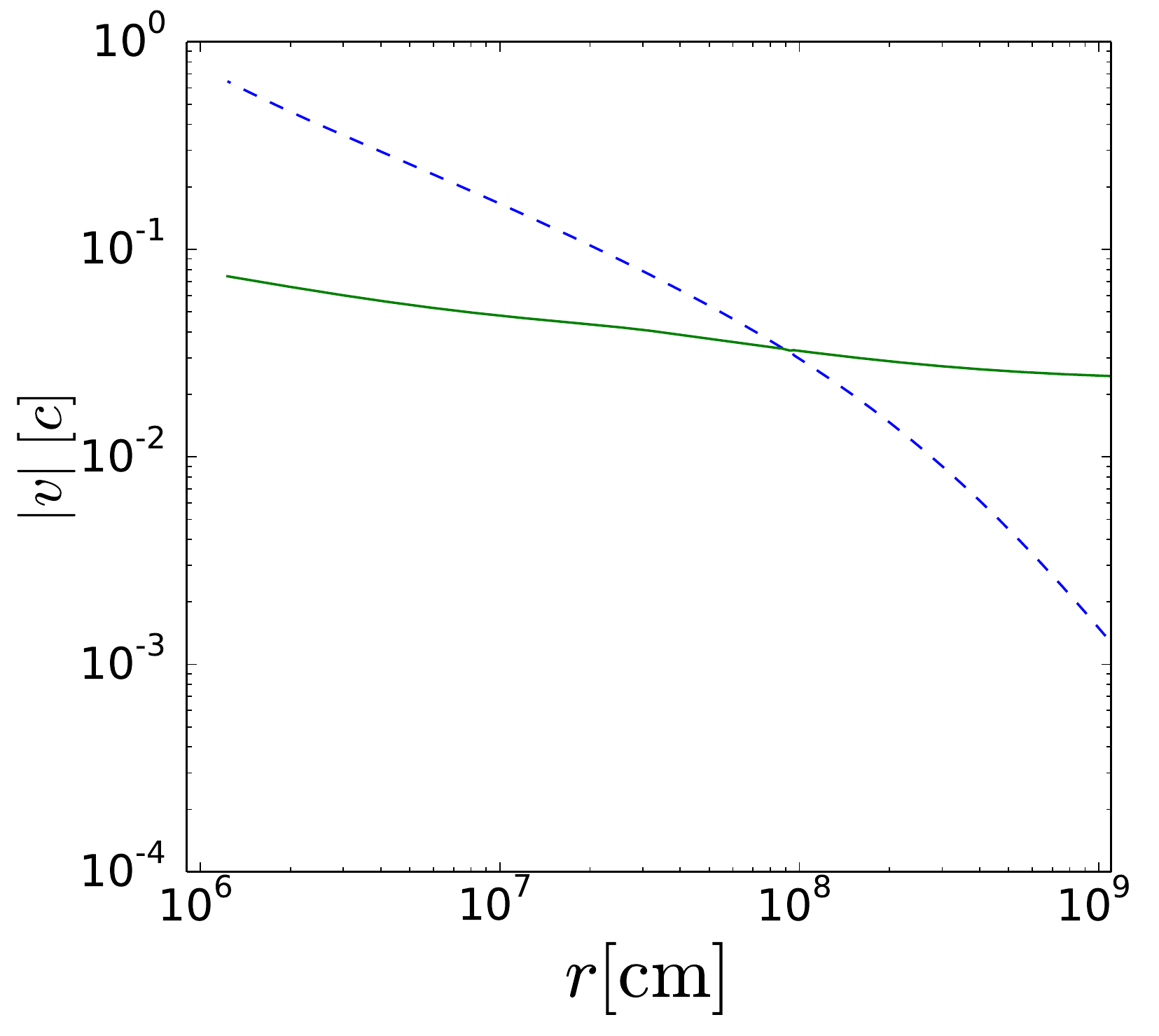}}
	\end{subfigure}
	\caption{Illustrative accretion solution for an accretion rate  $\dot{M}=10^{-5} M_{\odot}/s$ and entropy per baryon $s=80~ k_{\rm B}/{\rm nuc}$. The left panel shows the density (green-dashed line, left axis), pressure (red-solid line, right axis) and ram pressure (blue-dotted line, right axis). The right panel shows the absolute value of the fluid velocity (blue-dashed line) and the sound speed (green-solid line). The two lines cross at the critical point.}
	\label{fig:michel_solution}
\end{figure*}

 \section {Stationary spherical accretion}
\label{sec:michel}

We model the fallback of the reverse shock as the spherically symmetric accretion of an unmagnetized relativistic fluid. The stationary solutions for this system were first obtained by \cite{Michel:1972} for the case of a polytropic EoS. Here, we extend this work to account for a general (microphysically motivated) EoS.
 The equations that describe the motion of matter captured by a compact object, i.e.~a NS or black hole, can be derived directly from the equations of relativistic hydrodynamics , namely the conservation of rest mass,
 \begin{eqnarray}
 \nabla_{\mu}J^\mu=0\,,
 \end{eqnarray}
and the conservation of energy-momentum,
\begin{eqnarray}
 \nabla_\mu T^{\mu \nu}=0\,,
 \end{eqnarray}
where we use the notation $\nabla_{\mu}$  for the covariant derivative and  the density current $J^{\mu}$ and the (perfect fluid) energy-momentum tensor $T^{\mu\nu}$ are given by
\begin{eqnarray}
\label{eq:density_current}
J^{\mu} &=& \rho u^{\mu}\,,
\\
T_{\mu\nu} &=& \rho h u_{\mu}u_{\nu} + p g_{\mu \nu}\,.
\end{eqnarray}
In the above equations $\rho$ is the rest-mass density, $p$ is the pressure and $h$ is the specific enthalpy, defined by  $h = 1 + \varepsilon + p/\rho$, where  $\varepsilon$ is the specific internal energy, $u^{\mu}$ is the four-velocity of the fluid and  $g_{\mu \nu}$ defines the metric of the general spacetime where the fluid evolves.  Assuming spherical symmetry and a steady state we have
\begin{eqnarray}
\frac{d}{dr}(J^1\sqrt{-g} )&=& 0\,,
\label{eq:der_r_mass}
\\
\frac{d}{dr}(T_0^1\sqrt{-g}) &=& 0\,,
\label{eq:der_r_mom}
\end{eqnarray}
where $g\equiv \det(g_{\mu\nu})$. The exterior metric of a non-rotating compact object is given by the Schwarzschild metric 
\begin{eqnarray}
ds^2 &=& -\left(1-\frac{2{M}}{r}\right)dt^2+\left(1-\frac{2{M}}{r}\right)^{-1}dr^2 
\nonumber \\
&+& r^2 (d\theta^2+ \sin^2\theta \,d\varphi^2)\,.\label{eq:schwarzschild}
\end{eqnarray}
In Schwarzschild coordinates Eqs.~(\ref{eq:der_r_mass}) and (\ref{eq:der_r_mom}) can be easily integrated to obtain \citep[cf.][]{Michel:1972}
\begin{eqnarray}
\rho \, u \, r^2 &=& C_1\,, \label{eq:michel1}
\\
h^2 \left(1-\frac{2{M}}{r}+u^2\right)&=&C_2\, \label{eq:michel2}
\end{eqnarray}
where $C_1$ and $C_2$ are integration constants and $u\equiv u^r$. 
To obtain an adiabatic solution for the accreting fluid, we differentiate Eqs.~(\ref{eq:michel1}) and (\ref{eq:michel2}) at constant entropy and eliminate
$d\rho$ 
\begin{eqnarray}
\label{eq:crit_point}
&&\frac{du}{u}\left[ V^2-u^2 \left(1-\frac{2{M}}{r}+u^2   \right)^{-1}\right] 
\nonumber \\
&+& \frac{dr}{r}\left[ 2V^2-\frac{{M}}{r} \left(1-\frac{2{M}}{r}+u^2   \right)^{-1}\right] =0\,,
\end{eqnarray}
where 
\begin{eqnarray}
V^2\equiv\frac{\rho}{h}\left. \frac{\partial h}{\partial \rho} \right|_s\,.
\end{eqnarray}
The solutions of this equation are those passing through a critical point where both terms in brackets in equation~(\ref{eq:crit_point}) are zero,
i.e. those fulfilling
\begin{eqnarray}
2u_c^2 &=& \frac{{M}}{r_c}\,,
\nonumber \\
V_c^2 &=& u_c^2(1-3u_c^2)^{-1}\,, \label{eq:crit_point2}
\end{eqnarray}
where sub-index $c$ indicates quantities evaluated at the critical point.
The critical point can be identified as the sonic point, i.e. the point where the velocity of the fluid equals its own sound speed.  After some algebra,
it can be shown that the constant $C_1$ in equation~(\ref{eq:michel1}) is related to the accretion rate $\dot{M}$ by
\begin{eqnarray}
\dot{M}=-4\pi C_1.
\end{eqnarray}
Thereby we can obtain the accretion solution by simply selecting the mass accretion rate and the specific entropy of the fluid, which fixes the two
constants $C_1$ and $C_2$. We note that, for each pair of values, the system~(\ref{eq:crit_point2}) has two solutions, although only one represents a physical accretion solution ($|u|\to 0$ at $r \to \infty$). In this case the fluid is supersonic for radii below the critical radius and subsonic above.
Figure \ref{fig:michel_solution} displays one illustrative accretion solution for a mass accretion rate  $\dot{M}=10^{-5} M_{\odot}/\rm{s}$ and entropy per baryon $s=80k_{\rm B}/{\rm nuc}$. 

For the accreting material, we use the tabulated Helmholtz EoS \citep{Timmes:2000}, which is an accurate interpolation of the Helmholtz free-energy of the Timmes EoS \citep{Timmes_Arnett:1999}. Timmes EoS, and Helmholtz EoS by extension, include the contributions from ionized nuclei, electrons, positrons and radiation. By default, Timmes EoS uses the rest mass density $\rho\ [\rm{g}/\rm{cm}^3]$, temperature $T\ [\rm{K}]$ and composition as input. For convenience, we have developed a search algorithm that allows to call the EoS with different thermodynamical variables as input (e.g.  $\rho$, $s$ and composition  as inputs for the adiabatic flow of accreting material). Helmholtz EoS also requires the mean mass number $\bar{A}$ and the mean atomic number  $\bar{Z}$. 

At low densities, $\rho < 6\times 10^7$~g~cm$^{-3}$, and temperatures, $T\lesssim 2\times 10^9$~K, nuclear reactions proceed much slower than the accretion timescale and the composition remains frozen during the accretion. We fix the composition to that at the reverse shock formation point. Given the uncertainties, we consider two possibilities in this regime, either pure helium or pure carbon. At temperatures $T\gtrsim 2\times 10^9$~K nuclear burning becomes fast enough to change the composition. For $T\gtrsim 4\times 10^9$~K the fluid reaches nuclear statistical equilibrium (NSE) on a  significantly shorter  timescale than the accretion timescale \citep[see e.g.][]{Woosley:2002}. To deal with the high temperature regime, $T\ge 2\times 10^9$~K, we have tried three
different approaches: 1) unchanged composition of the accreting material, 2) compute the NSE composition at a given temperature and density using a thermonuclear reaction
network with 47 isotopes \citep{Timmes:1999, Seitenzahl:2008} and 3) simplified burning with four transitions:  $^{4}\rm{He}$ for $T\le 2\times 10^9$~K, $^{56}\rm{Ni}$ for $2\times 10^9 > T\ge 5\times 10^9$~K, $^{4}\rm{He}$ for $5\times 10^9 > T\ge 2\times 10^{10}$~K and protons and neutrons for $T> 2\times 10^{10}$~K. We use the publicly available routines of the Hemlholtz EoS and the NSE equilibrium kindly provided by the authors\footnote{http://cococubed.asu.edu/code\_pages/codes.shtml}.

\section{Non-magnetized accretion and pile-up}
\label{sec:non-magnetized accretion}

Before considering the case of magnetized accretion onto a NS, we study the case of non-magnetized accretion. For the span of accretion rates considered in this work, the sonic point of the accreted fluid is located at $r>23500$ km at entropy $s=10~k_{\rm{B}}/\rm{nuc}$, and hence
the fallback material falls supersonically onto the NS. Inevitably an accretion shock forms at the surface of the star, which propagates outwards. The accreted
fluid crossing the shock will heat up, increasing its specific entropy and will fall subsonically. The high entropy of this material ($s_{shock} \in [70 - 300]~k_{\rm B}/{\rm nuc}$) and the compression that experiments as it flows inwards raises the temperature beyond the pair creation threshold, $T_{\rm pair}\approx 10^{10}$~K,
and the fluid will cool efficiently via neutrino-antineutrino annihilation. Therefore, the kinetic energy of the supersonically accreting fluid is mostly transformed into
thermal energy as it crosses the accretion shock and then is dissipated to neutrinos close to the NS surface. \cite{Chevalier:1989} showed that the accretion shock
will eventually stall at a certain radius as an energy balance is found. The radius of the stalled shock depends only on the accretion rate $\dot M$
and is located at about $R_{\rm shock} \sim 10^7 - 10^8$~km. In some estimates bellow in this work we use the values provided in table~1 in \cite{Houck:1991}, based in
a more realistic treatment of the accretion and neutrino cooling.

\begin{figure}
  \centering
  {\includegraphics[width=0.5\textwidth]{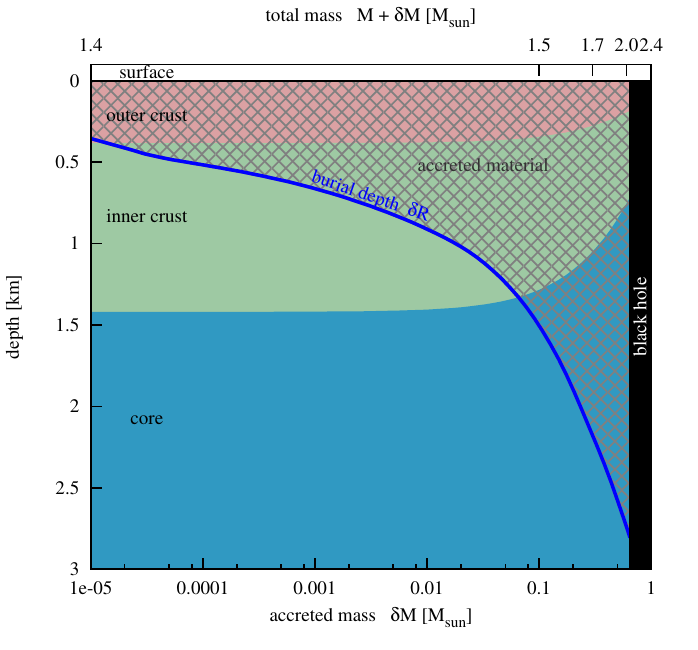}}
  \caption{Dependence of the burial depth (blue solid lines), $\delta R$, with the accreted mass, $\delta M$ (bottom axis), for a $M=1.4 M_\odot$ NS using
    APRDH EoS. Note, that positive values of the depth increase downwards. Regions occupied by the outer crust, inner crust and core appear with different
    colors and labeled. The region occupied by the accreted material is
    plotted  with a gray crosshatch pattern. The top axis shows the total mass of the NS after accretion, $M+\delta M$. Above $M=2.25~M_\odot$ the configuration
    is unstable and the object will collapse to a black hole.
   }
  \label{fig:drdm}
\end{figure}

The final fate of the neutrino-cooled material falling steadily onto the NS surface is to pile up on top of the original NS material forming
a layer of new material. In order to study the effect of the pile up we consider a NS of mass $M$ and radius $R$. If we add a mass $\delta M$
to the equilibrium model, the new NS will have a new radius $R_{\rm new}$ smaller than the original one. The original surface of the star, will now be buried
at a depth $\delta R$, i.e. the new surface will be located at a distance $\delta R$ over the old surface. Although trivial, the last statement is important because
most of the discussion below in this work is carried out in terms of $\delta R$ and in terms of distances with respect to the original NS surface.
Therefore it makes sense to try to compute what is the dependence of the burial depth, $\delta R$, with the total accreted mass, $\delta M$. In order to compute this
dependence we use the TOV equations to solve a sequence of NS equilibrium models starting with $M$ and progressively increasing to $M+\delta M$ for different
values of $\delta M$. For each model in the sequence we compute $\delta R$ as the distance between the radius enclosing a mass $M$ and the surface of the star, i.e. the radius enclosing $M + \delta M$. Given the small values of $\delta M$, we integrate the TOV equations using a simple forward Euler method, with a step limited to relative variations
of density of $10^{-5}$ and a maximum step of $10$~cm. We have computed the relation between $\delta R$ and $\delta M$ for four different NS masses, $M=1.2$, $1.4$, $1.6$, $1.8$ and $2.0 M_\odot$.
We have used several realistic EoS in tabulated form, namely four different combinations
using either EoS APR \citep{Akmal:1998} or EoS L \citep{Pandharipande:1995} for the core and EoS NV \citep{Negele:1973} or EoS DH \citep{Douchin:2001}
for the crust. For each case we compute the sequence up to the maximum mass; beyond that mass, the equilibrium model is unstable and it will collapse to a
black hole in dynamical timescales. All EoS allow for equilibrium solutions with maximum mass consistent with recent observations of a NS with mass close to $2 M_\odot$ \citep{Demorest:2010,Antoniadis:2013}. The blue solid line in Fig.~\ref{fig:drdm} shows the dependence of $\delta R$ with $\delta M$ for a $1.4 M_\odot$ NS with the APRDH EoS. All other EoS and
neutron star masses show similar behavior. For all EoS, any amount of accreted mass larger than $\sim 10^{-4} M_{\odot}$ will sink the original NS surface to
the inner crust, and for $\delta M \sim 0.1 M_{\odot}$ the entire crust is formed by newly accreted material. The bottom line is that, if the accreted material
is able to compress the magnetosphere and deposit itself on top of the NS, the magnetic field trapped with the fluid may be buried into the NS crust,
and depending on the conditions (accreted mass and magnetic field strength), the burial depth could be as deep as the inner crust. We study next the impact of
magnetic fields in the vicinity of the NS, namely the magnetosphere, in the burial process.

\section{Magnetosphere}
\label{sec:magnetosphere}

\subsection{Potential magnetospheric solution}
\label{sec:forcefree}

For simplicity in the following discussion we use a {\it reference model} with the APRDH EoS and $M=1.4 M_\odot$. This model results in a
a NS with coordinate radius $R=12.25$ km. The effect of the EoS and the NS mass are discussed later in the text.
Given that both the magnetosphere and the accreted material involve low energy densities compared with those inside the NS, the spacetime
outside the NS can be regarded as non-self-gravitating and approximated by the Schwarzschild exterior solution. 

\begin{figure*}
	\begin{subfigure}
	{\includegraphics[width=55mm]{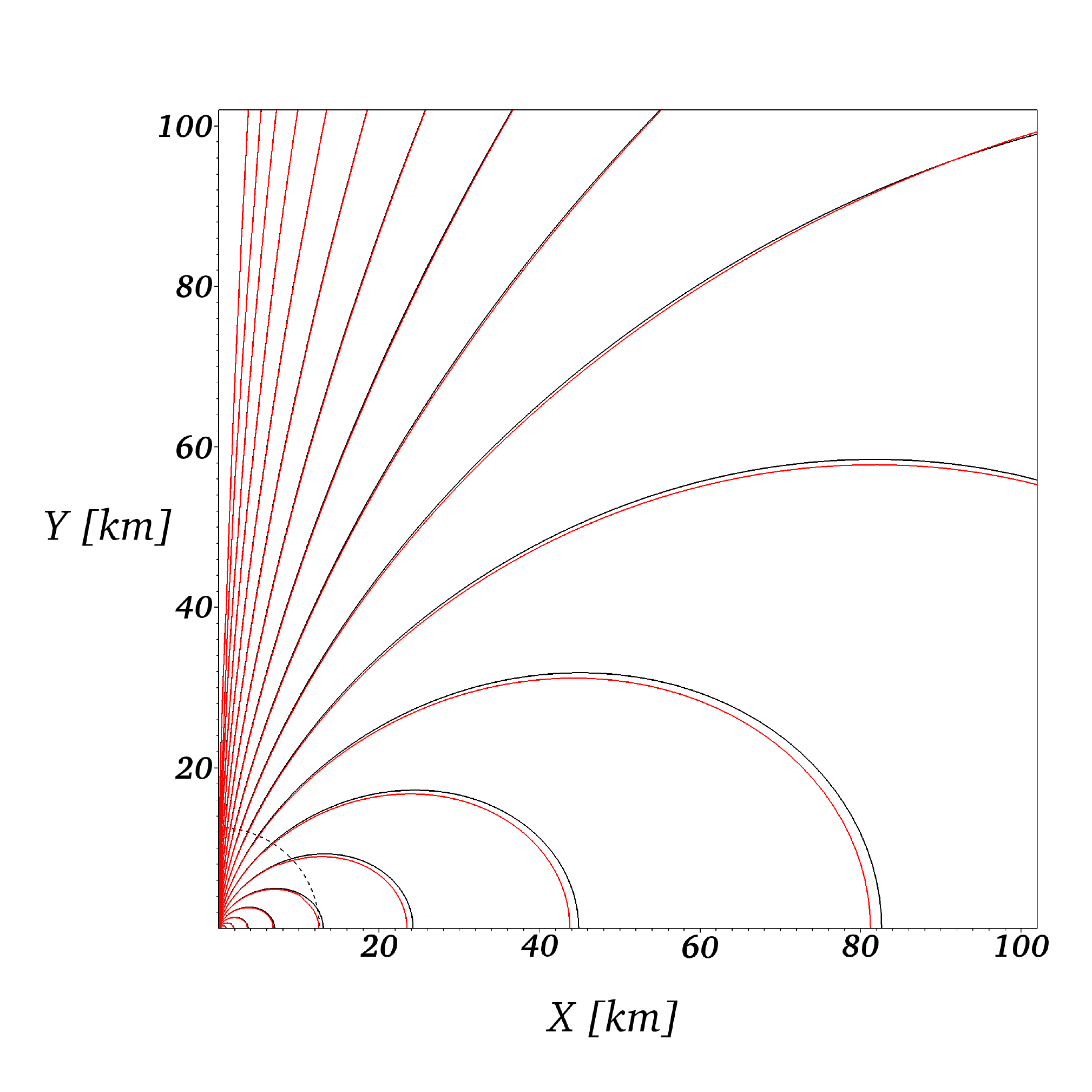}}
	\end{subfigure}
	\begin{subfigure}
	{\includegraphics[width=55mm]{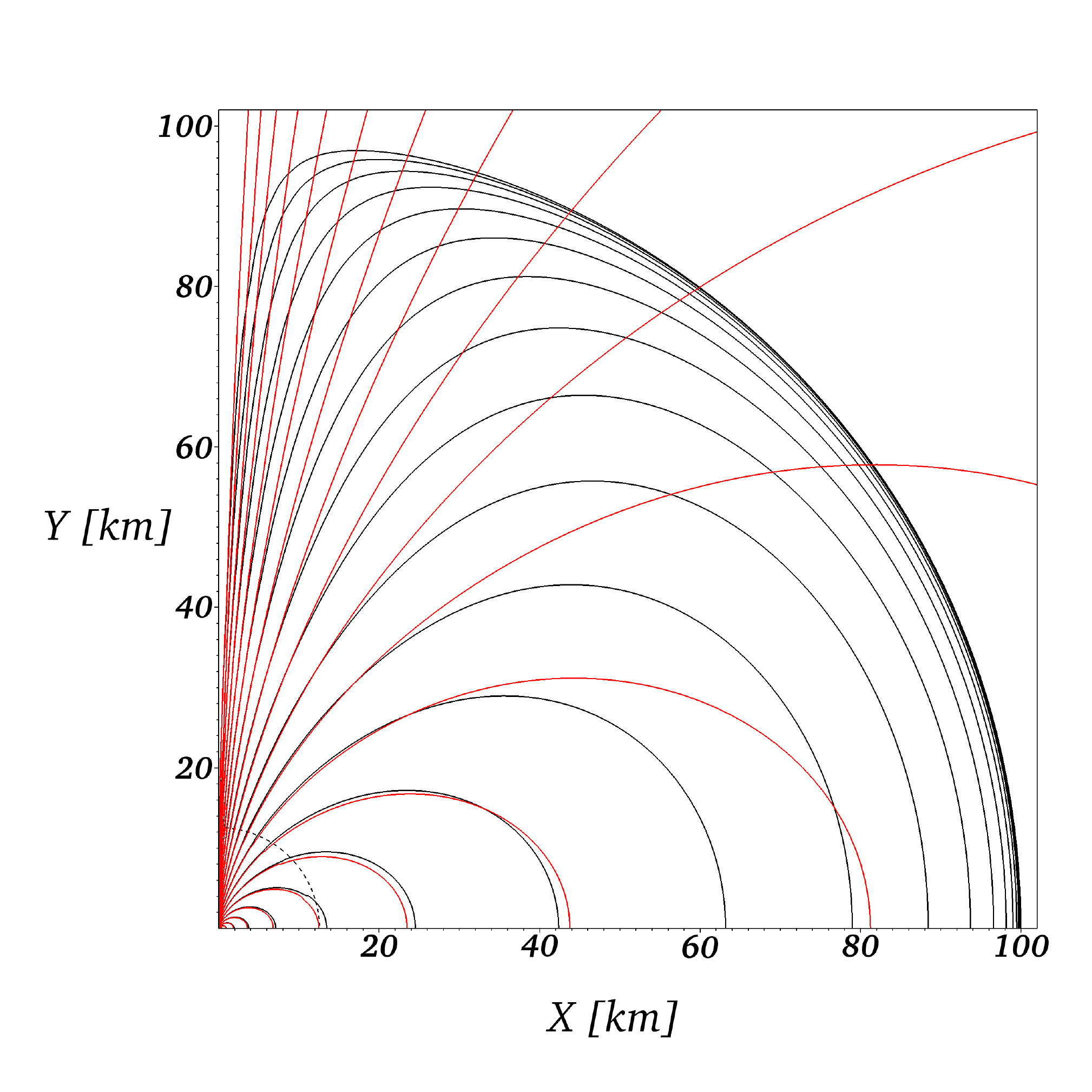}}
	\end{subfigure}
	\begin{subfigure}
	{\includegraphics[width=55mm]{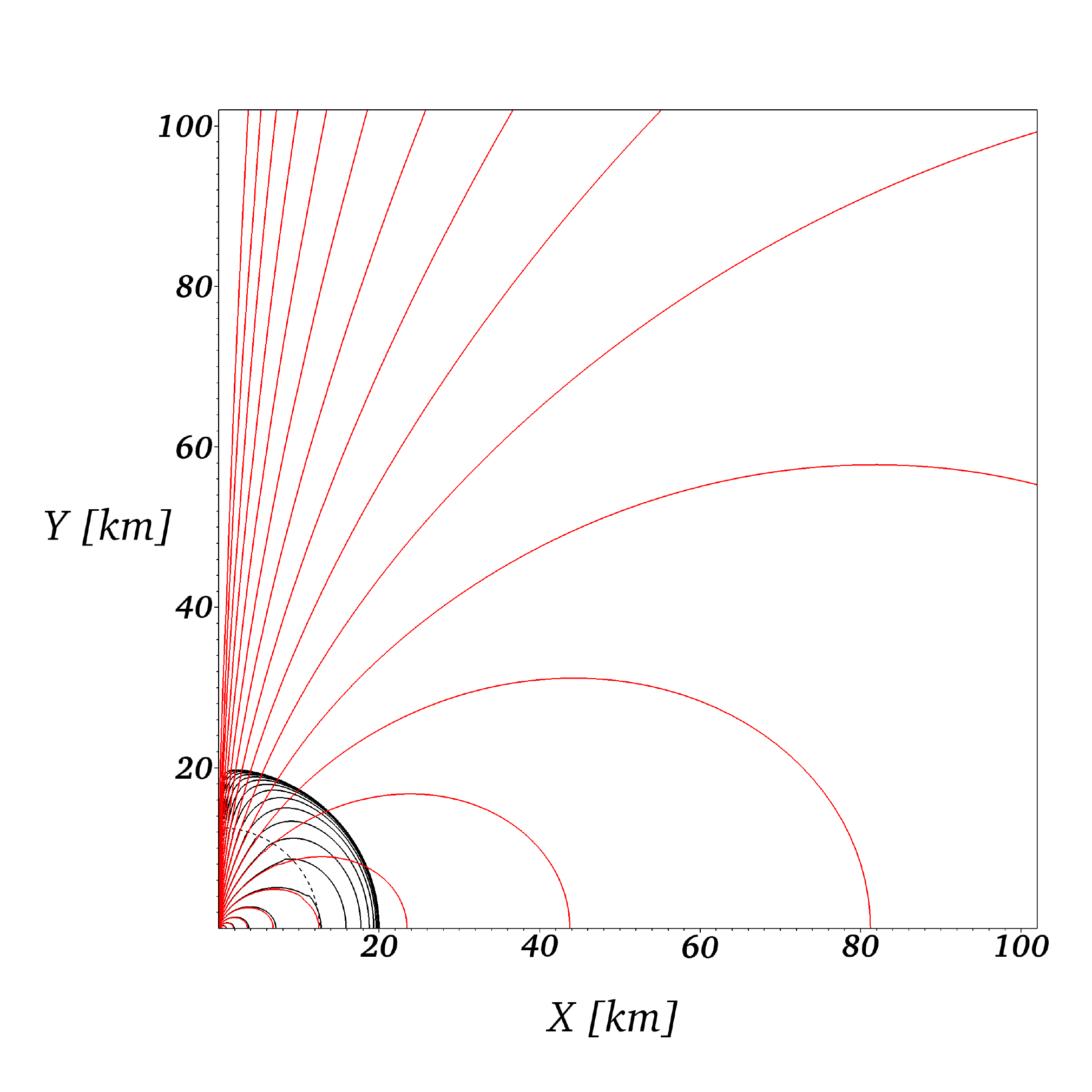}}
	\end{subfigure}
	\vspace{-5mm}
	\caption{Magnetic field lines (black lines) for three different positions of the magnetopause for the same initial distribution of the magnetic field (red lines) with $B_p = 10^{13}$~G. The dashed line represents the NS surface.}
	\label{mf_compression}
\end{figure*}

The magnetosphere extends between the NS surface and the magnetopause, which will be assumed to be a spherically symmetric surface
at the location of the  infaling reverse shock. We model this region using the force-free magnetic field approximation, $ \mathbf{J}\times\mathbf{B}=0$,  $\mathbf{J}$ being
the electric current and $\mathbf{B}$ the magnetic field. We neglect the currents resulting from the rotation of the star.
Consequently the magnetic field has a potential solution, solution of the relativistic
Grad-Shafranov equation. In spherical coordinates, the magnetic field vector components are related to the vector potencial $\mathbf{A}$ as,
\begin{eqnarray}
\label{Br}
\hat{B}_r&=&\frac{1}{r^2\sin \theta}\partial_\theta A_\phi \,,
\label{Br}
\\
\hat{B}_\theta&=&\frac{-1}{r^2\sin \theta}\partial_r A_\phi \,,
\label{Btheta}
\\
\hat{B}_\phi &=& 0\,,
\end{eqnarray}
where $\hat{B}_i=\sqrt{\gamma}B_i$ and $\gamma$ is the determinant of the spatial metric.
If we assume axisymmetry, the unique nonzero component of the electric current is the $\phi$ component,
\begin{eqnarray}
  J_\phi = \sin \theta \left[ \partial_r(r \hat{B}_\theta)-\partial_\theta \hat{B}_r\right].
  \label{eq:current}
\end{eqnarray}
Imposing the force-free condition, we obtain,
\begin{eqnarray}
-J_\phi \hat{B}_\theta &=& 0\,,
\\
J_\phi \hat{B}_r &=& 0\,.
\end{eqnarray}
Since $\hat{B}_r, \hat{B}_\theta\neq0$, the only possible solution is $J_\phi=0$. 
As we want an expression that only depends on the vector potential, we replace  Eqs.~(\ref{Br}) and (\ref{Btheta}) in equation~(\ref{eq:current}) resulting in 
\begin{eqnarray}
\label{jphi}
J_\phi & =&\sin \theta \left[\frac{-1}{\sin \theta} \partial_r(\partial_r)A_\phi-\frac{1}{r^2}\partial_\theta \left(\frac{\partial_\theta A_\phi}{\sin\theta}\right)\right] 
\nonumber \\
&=&-\partial_{rr}A_\phi-\frac{1}{r^2}\partial_{\theta\theta}A_\phi+\frac{\cot \theta}{r^2}\partial_\theta A_\phi=0\,.
\end{eqnarray}
We discretize this expression using second order finite differences and solve the resulting linear system of equations using a cyclyc reduction
algorithm \citep{Swarztrauber:1974}. We impose Dirichlet boundary conditions on $A_\phi$ at the surface of the NS to match with the interior
value of the radial component of the magnetic field. Our aim is to describe a magnetosphere, which is confined within a certain radius, $R_{\rm mp}$,
defining the magnetopause. Magnetic field lines at the magnetopause are parallel to this interface and they enter the NS along the axis.
Therefore, they correspond to lines with $A_\phi =0$, which we use as Dirichlet boundary condition at $R_{\rm mp}$ to solve the Grad-Shafranov equation.
We can obtain the field distribution after the compression by simply changing the radius where the boundary conditions are imposed. The evolution of the magnetic field geometry before and after compression is shown in Fig.~\ref{mf_compression} for three illustrative cases.

For the interior magnetic field, which determines the boundary conditions at the surface of the star, we use two different magnetic field distributions, a dipolar magnetic field ({\it dipole} herafter) and a poloidal field generated by a circular loop of radius $r=4\times10^5$ cm \citep{Jackson:1962}
({\it loop current} hereafter). 
Following~\citet{Gabler:2012}, it is useful to introduce the
{\it equivalent magnetic field}, $B^*$, which we define as the magnetic field strength at the surface of a Newtonian, 
uniformly magnetized sphere with radius $10$~km
having the same dipole magnetic moment as the configuration we want to describe. It spans the range $B^*\in[10^{10}-10^{16}]$~G.

\subsection{Magnetosphere compression}
\label{sec:riemann}

In the case of a fluid accreting onto a force-free magnetosphere, the magnetopause will remain spherical and will move inwards as long as the total pressure of the unmagnetized fluid, $p_{\rm tot} = p + p_{\rm ram}$
, exceeds that of the magnetic pressure, $p_{\rm mag}$, of the magnetosphere. If we approximate the magnetopause as a spherical boundary between the spherically symmetric accreting solution described in Section~\ref{sec:michel} and the potential solution computed in Section~\ref{sec:forcefree}, its properties can be described as the solution of a Riemann problem at the magnetopause. Since the magnetic field of the initial state is tangential to the magnetopause, we can use the exact solution of the Riemann problem developed by \cite{Romero:2005}. A succinct summary of the details of the implementation of the Riemann solver can be found in Appendix~\ref{riemann_problem}.

\begin{center}
	\begin{figure*}
		{\includegraphics[width=80mm]{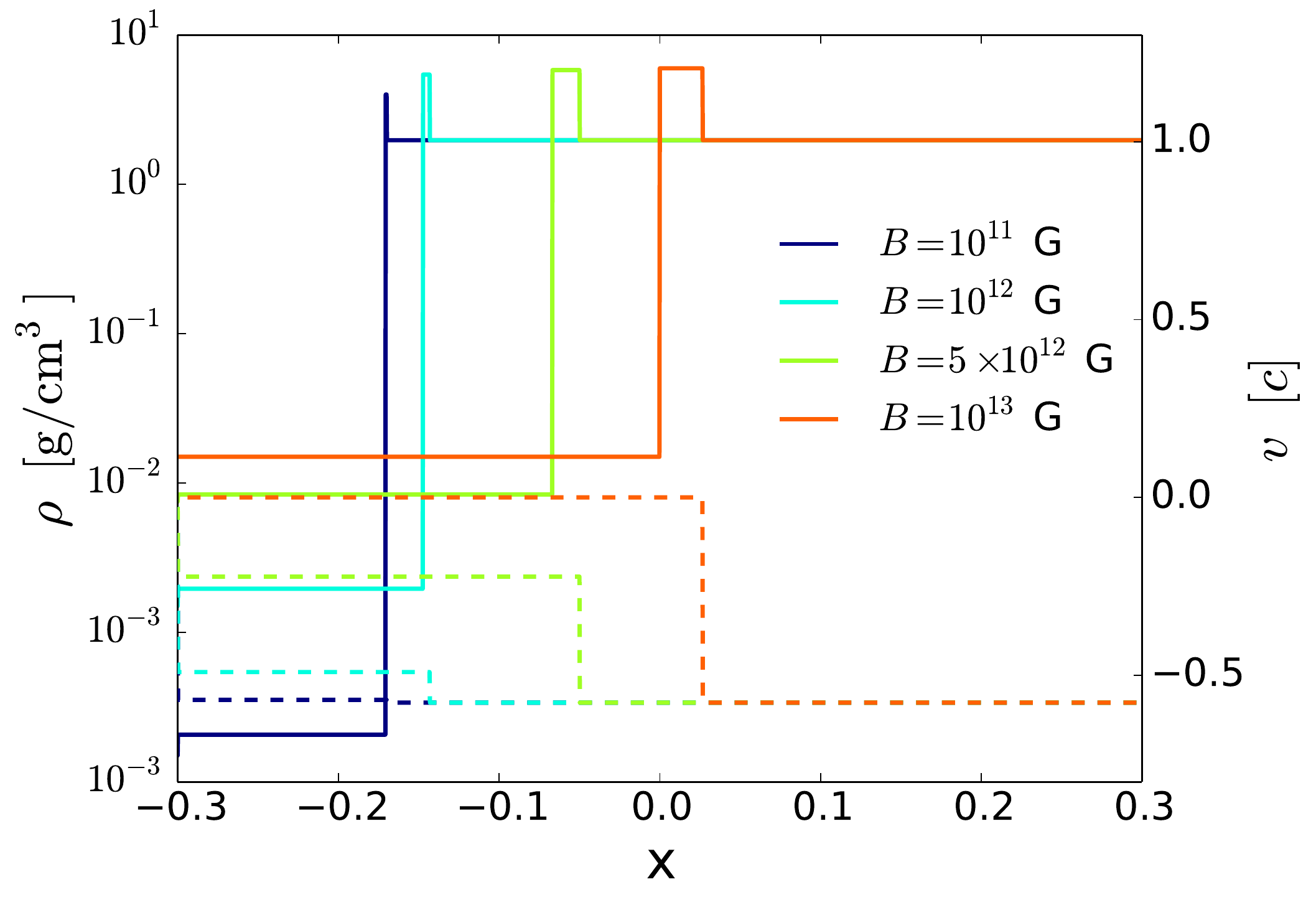}}
		{\includegraphics[width=80mm]{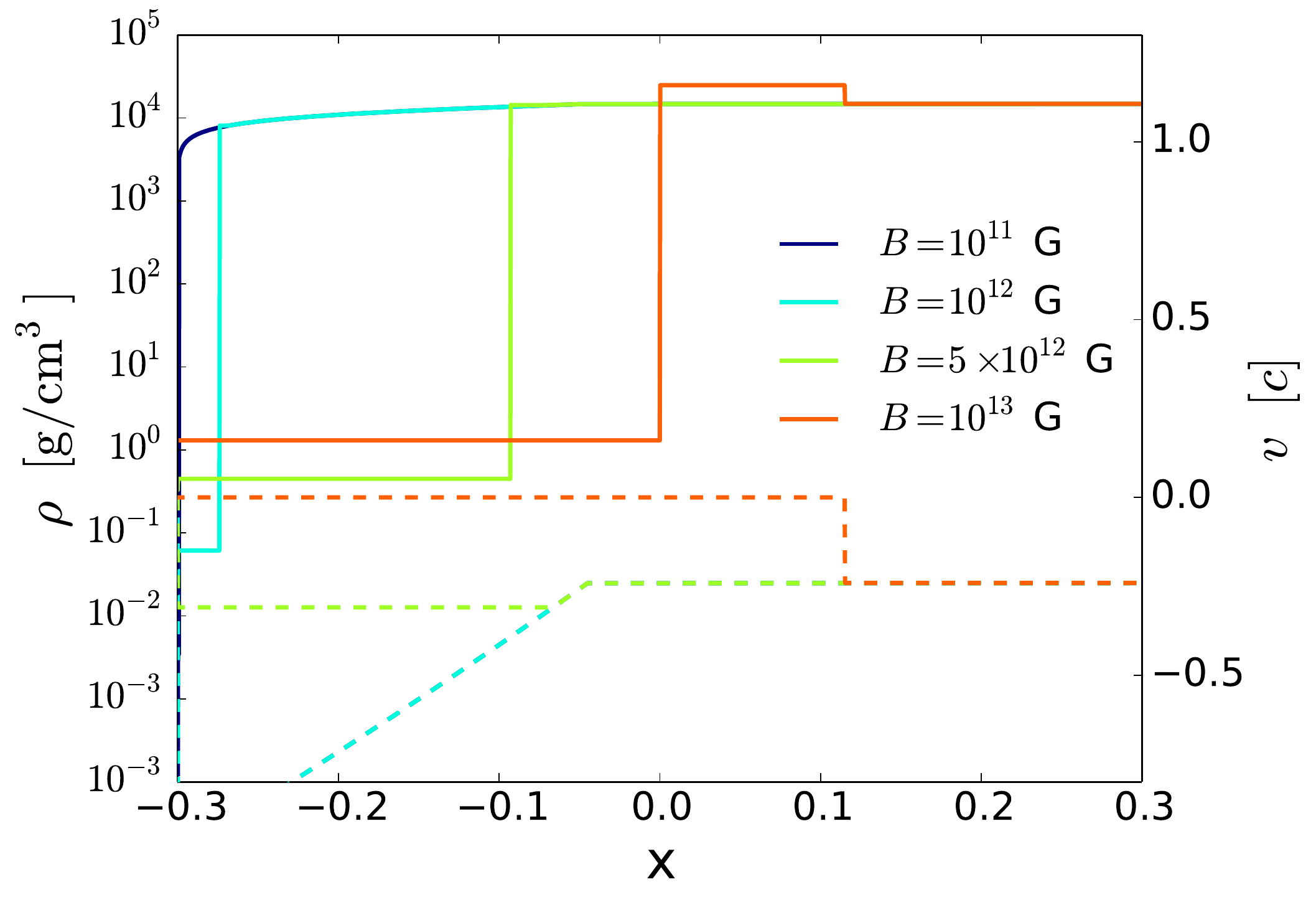}}
		\caption{Density (solid lines, left axis) and velocity (dashed lines, right axis) profiles of the solution of the Riemann problem for several values of the magnetic field. Initially the discontinuity is set at $x=0$, an accreting fluid at $x>0$ and a magnetized fluid at $x<0$, with constant magnetic field $B$. The left panel shows the case of supersonic accretion of a fluid with specific entropy $s=10k_{\rm{B}}/{\rm nuc}$ and $\dot{M}=10^{-7}\ \rm{M}_\odot/\rm{s}$ at $t=0.3$ s . The right panel shows the case of subsonic accretion of a fluid with $s = 2000k_{\rm{B}}/{\rm nuc}$ and $\dot{M}=10^{-5}\ \rm{M}_\odot/\rm{s}$ at $t=0.3$ s.}
		\label{fig:riemann_problem}
	\end{figure*}
\end{center}

For illustrative purposes the left panel of Fig.~\ref{fig:riemann_problem} shows the solution of the Riemann problem for a supersonic fluid accreting from
the right into a magnetically dominated region (magnetosphere) on the left.  The figure displays both the density (left axis, solid lines) and the fluid velocity (right axis, dashed lines). The initial discontinuity is located at $x=0$. The right constant state of the Riemann problem corresponds to the accreting fluid with an entropy of $s=10~k_{\rm{B}}/\rm{nuc}$ and accretion rate of $\dot{M}=10^{-7} \ \rm{M}_\odot/\rm{s}$. The left constant state corresponds a state with magnetic pressure $B^2/2$. The figure plots the corresponding solutions for different values of $B$ around the equilibrium  (indicated in the legend).

Looking at the left panel of Fig.~\ref{fig:riemann_problem} from left to right, the first jump in density corresponds to the contact discontinuity, point at which, as expected, the velocity remains continuos. The next discontinuity is a shock wave, where both the density and velocity are discontinuous, and both decrease. For low magnetic fields, $B\le 10^{10} G$, the low magnetic pressure on the left state cannot counteract the total pressure of the accreting fluid and the contact discontinuity advances to the right at a velocity equal to that of the accreting fluid; a shock front is practically nonexistent. As the magnetic field is increased the velocity of the contact discontinuity decreases and it becomes zero at about $B=10^{13}$ G. We identify this point as the {\it equilibrium point}, since no net flux of matter crosses $x=0$. Around this equilibrium point, an accretion shock appears, which heats and decelerates matter coming from the right. The equilibrium point corresponds to a solution in which the matter crossing the shock has zero velocity, i.e. it piles
up on top of the left state as the shock progresses to the right.

The actual accretion of matter onto a magnetically dominated magnetosphere is expected to behave in a similar way as the described Riemann problem. At large
distances (low $B$) the magnetopause (contact discontinuity) is compressed at the speed of the fluid. As the magnetosphere is compressed, the magnetic
field strenght raises and at some point an equilibrium point is found, beyond which the magnetosphere impedes the accretion of the fluid.

In the right panel of Fig.~\ref{fig:riemann_problem} we show for the sake of completeness the solution for a subsonic accreting fluid. In accreting NS this regime
is probably unrealistic, since very large specific entropy is necessary ($s=2000 k_{\rm B}/{\rm nuc}$ in the example plotted). In this case the
solution is qualitatively different; instead of a shock, a rarefaction wave if formed for $B$ below the equilibrium point. For larger values of $B$,
an accretion shock is formed.
%

\begin{figure*}
	\begin{subfigure}
	{\includegraphics[height =40mm]{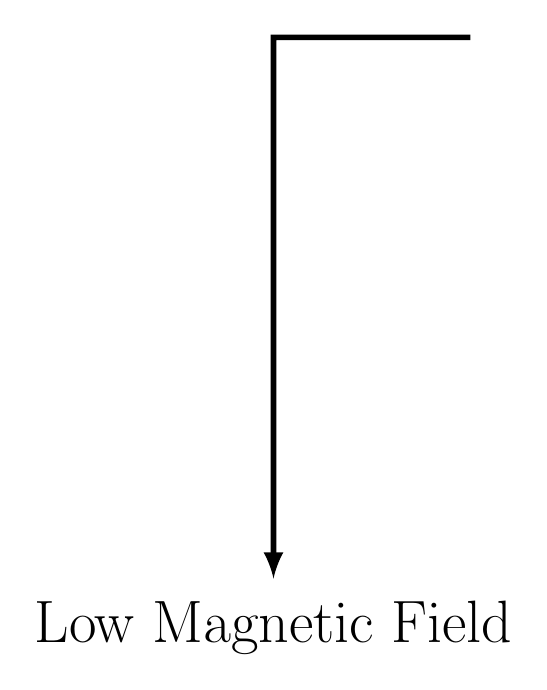}}
	\end{subfigure}
	\begin{subfigure}
	{\includegraphics[width=60mm]{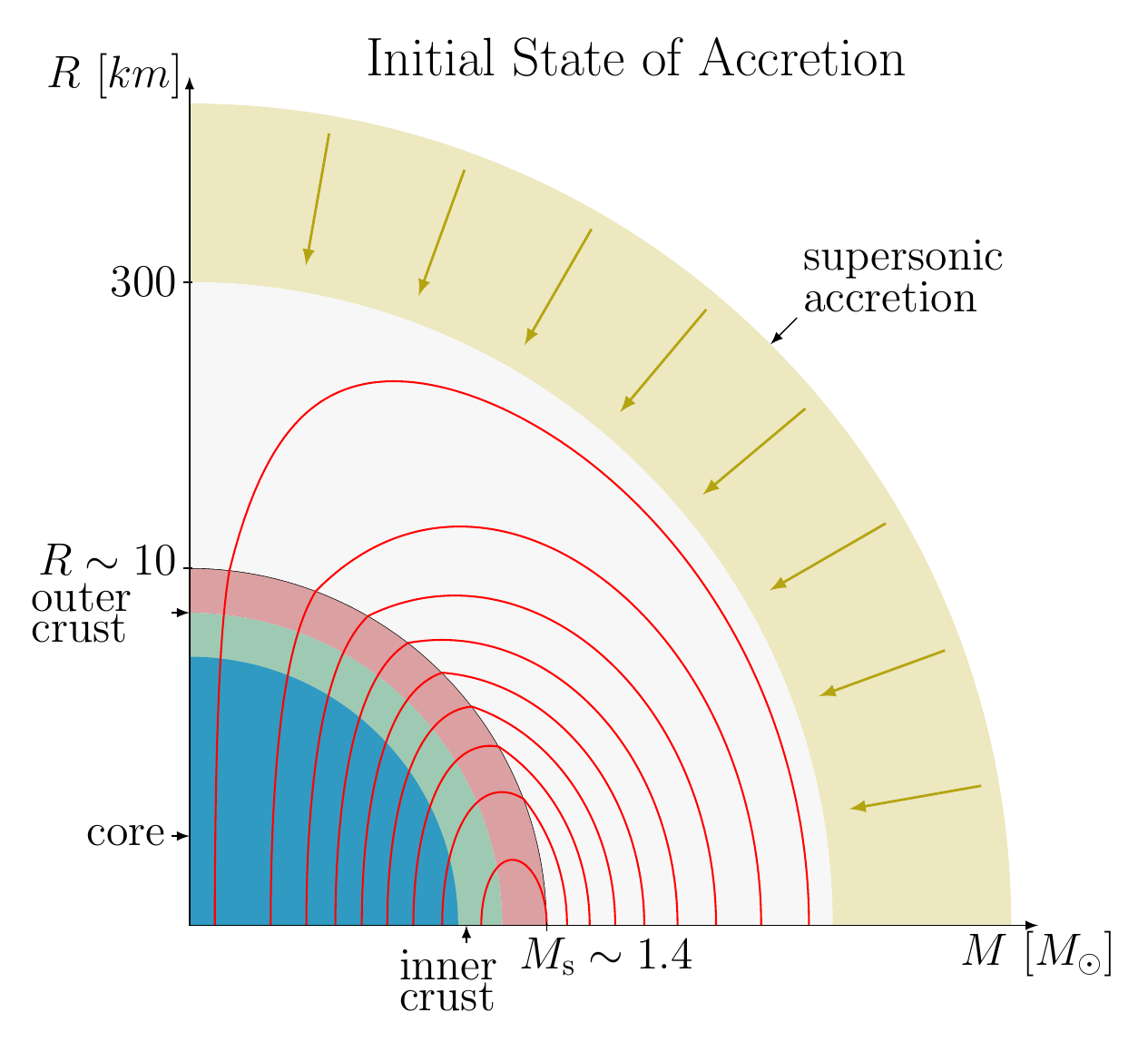}}
	\end{subfigure}
	\begin{subfigure}
	{\includegraphics[height =40mm]{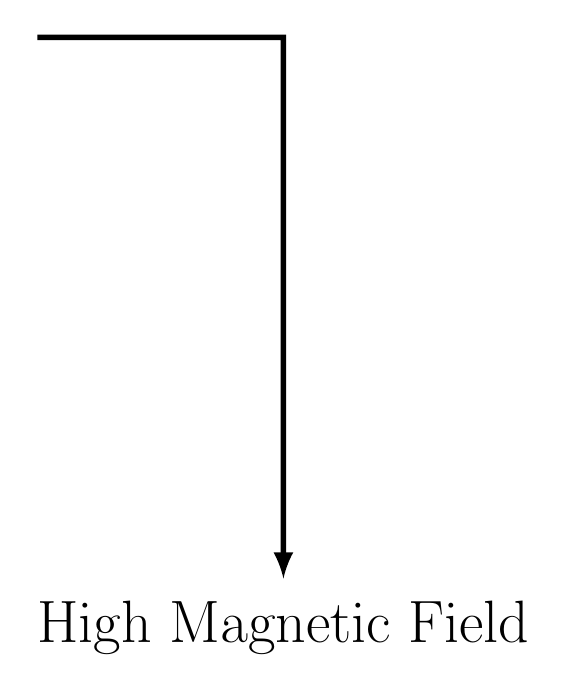}}
	\end{subfigure}
	\\
	\begin{subfigure}
	{\includegraphics[width=60mm]{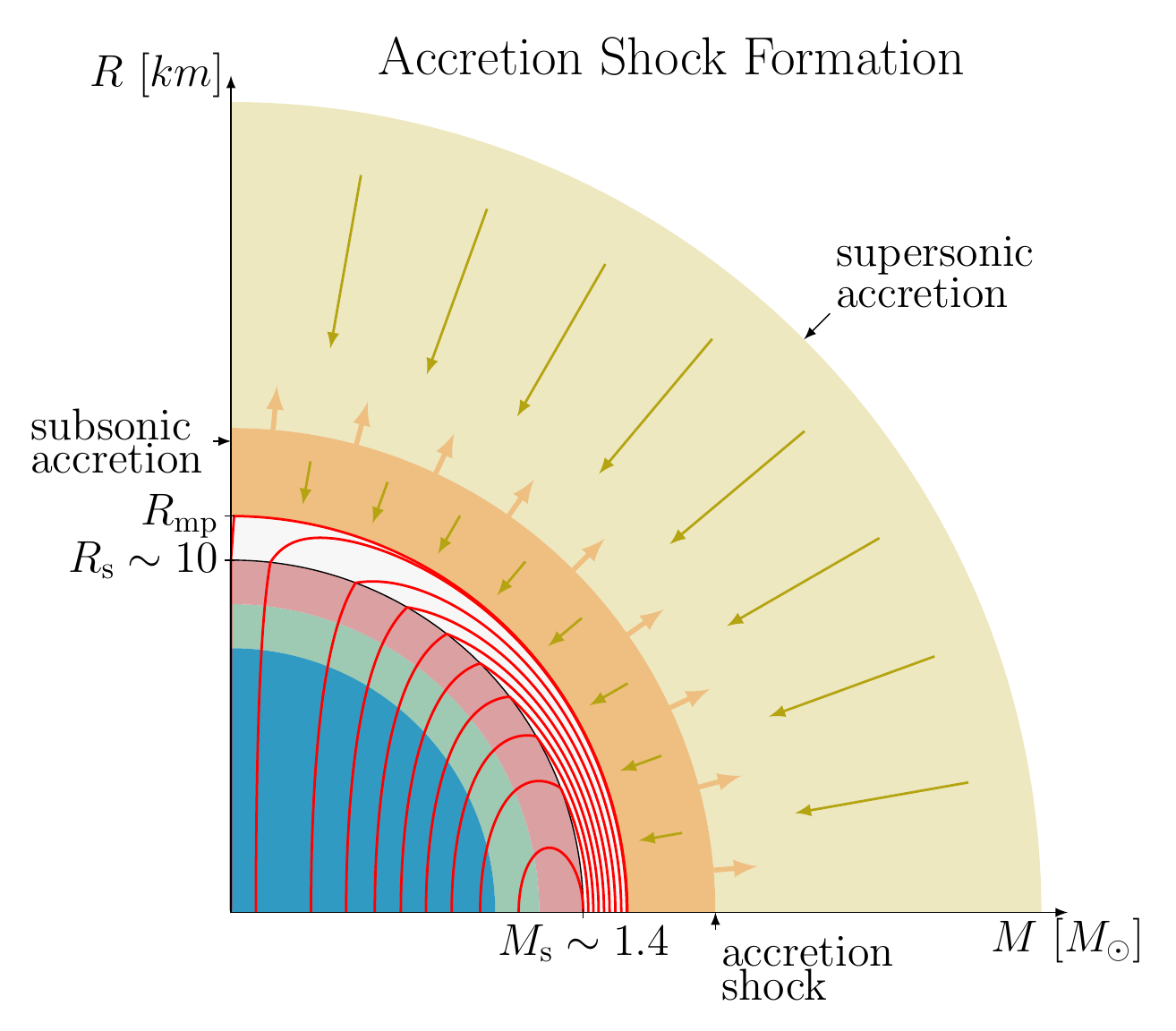}}
	\end{subfigure}
	\hspace{15mm}
	\begin{subfigure}
	{\includegraphics[width=60mm]{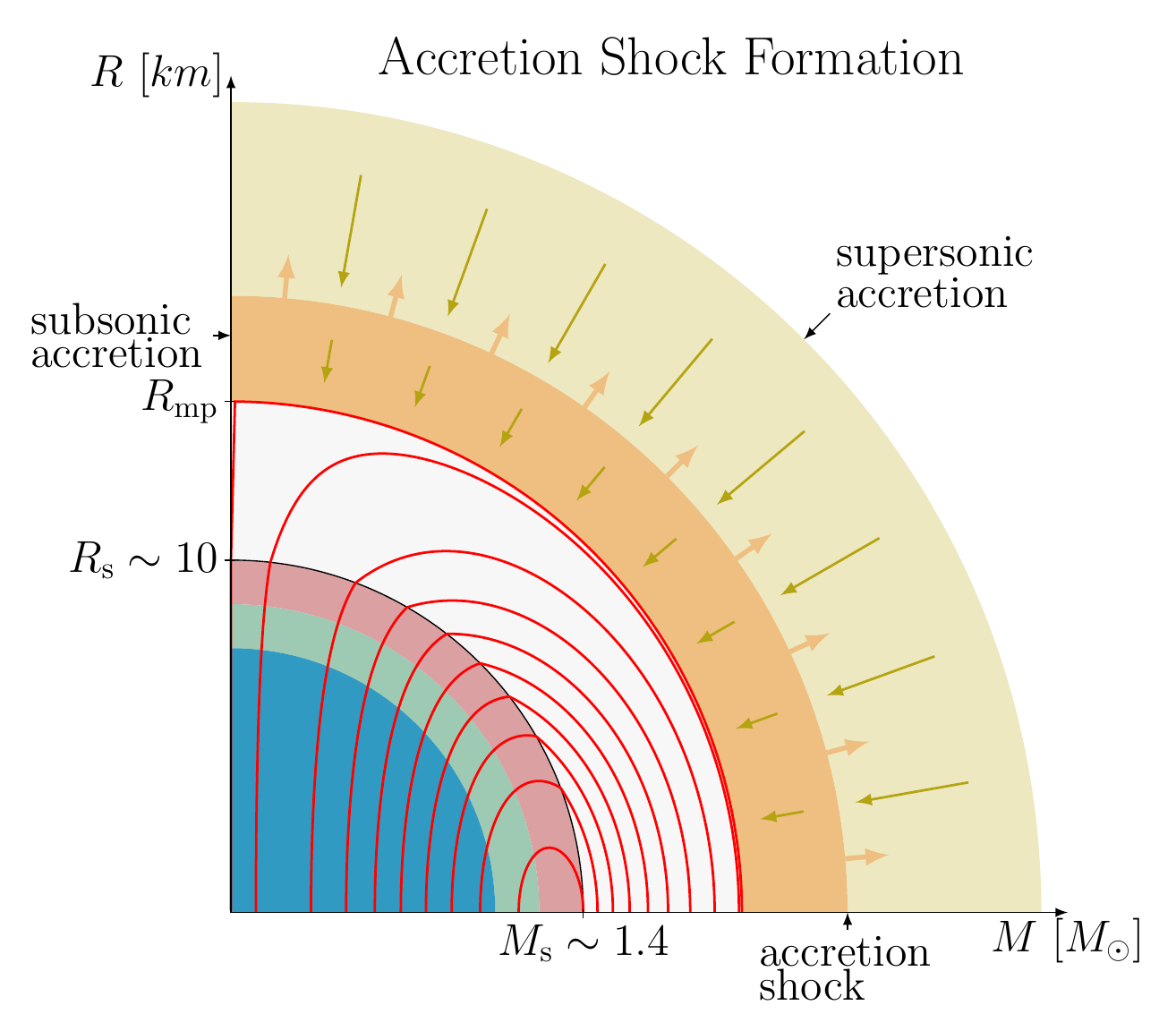}}
	\end{subfigure}
	\\
	\begin{subfigure}
	{\includegraphics[width=60mm]{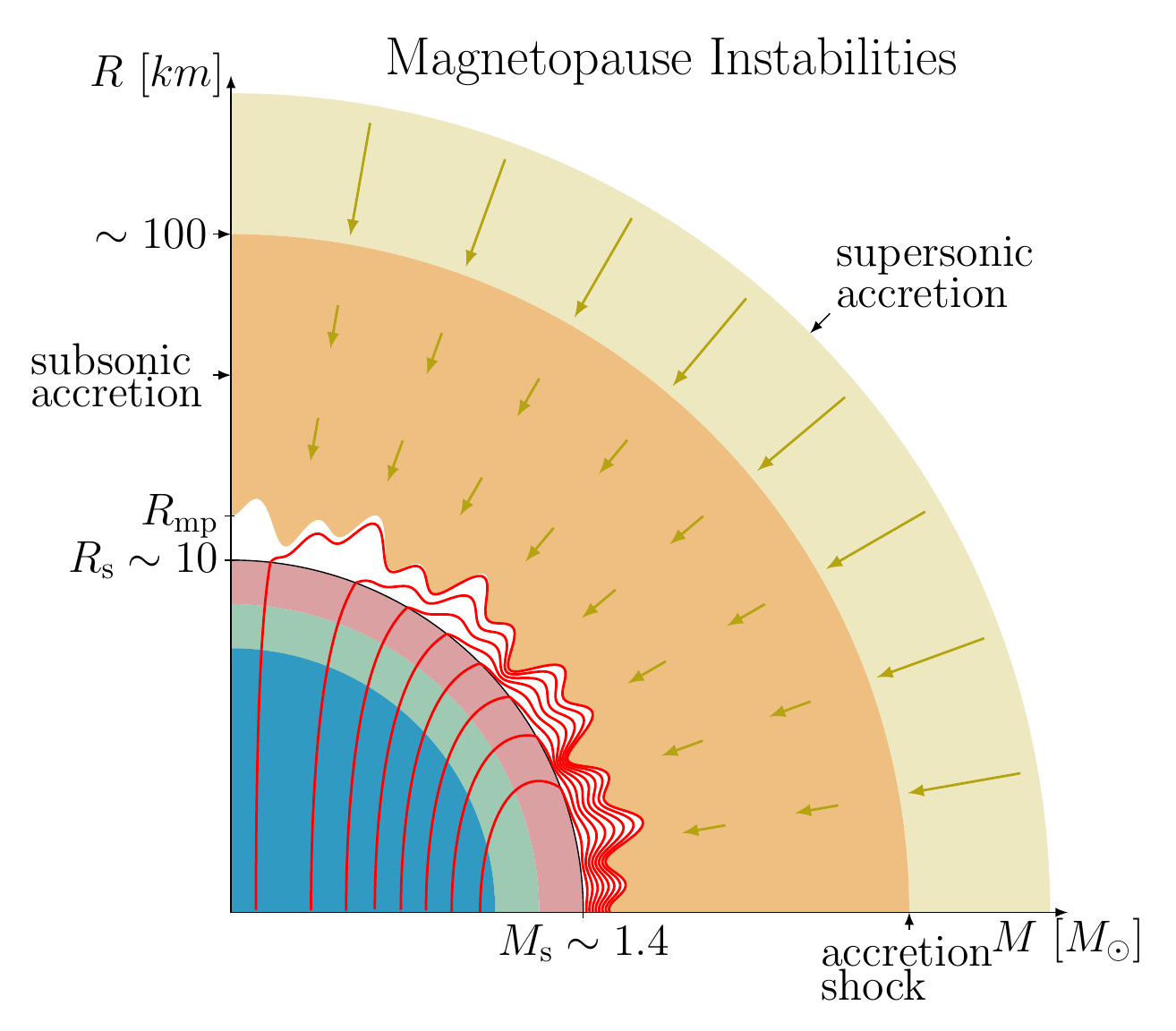}}
	\end{subfigure}
	\hspace{15mm}
	\begin{subfigure}
	{\includegraphics[width=60mm]{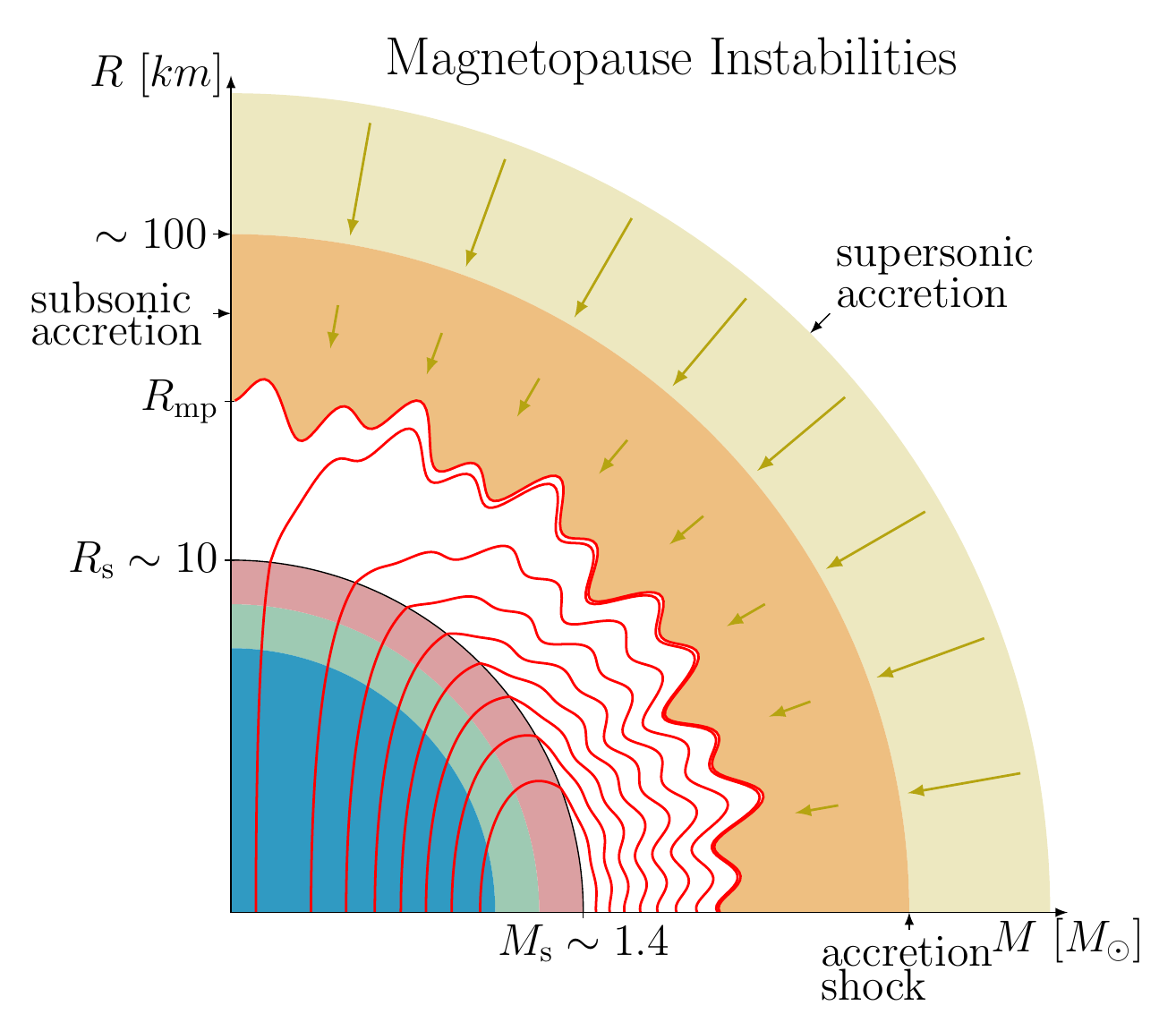}}
	\end{subfigure}
	\\
	\begin{subfigure}
	{\includegraphics[width=60mm]{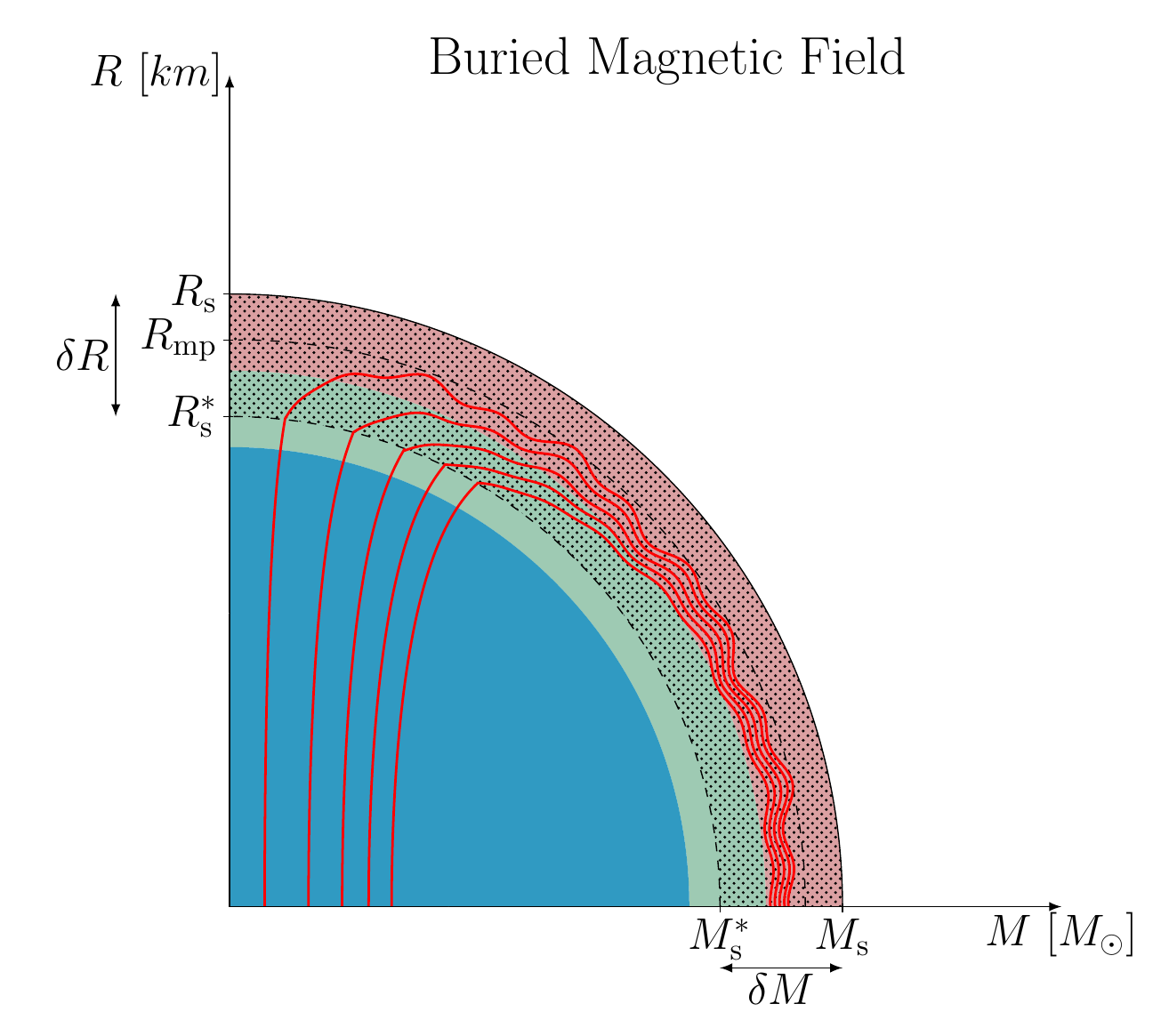}}
	\end{subfigure}	
	\hspace{15mm}
	\begin{subfigure}
	{\includegraphics[width=60mm]{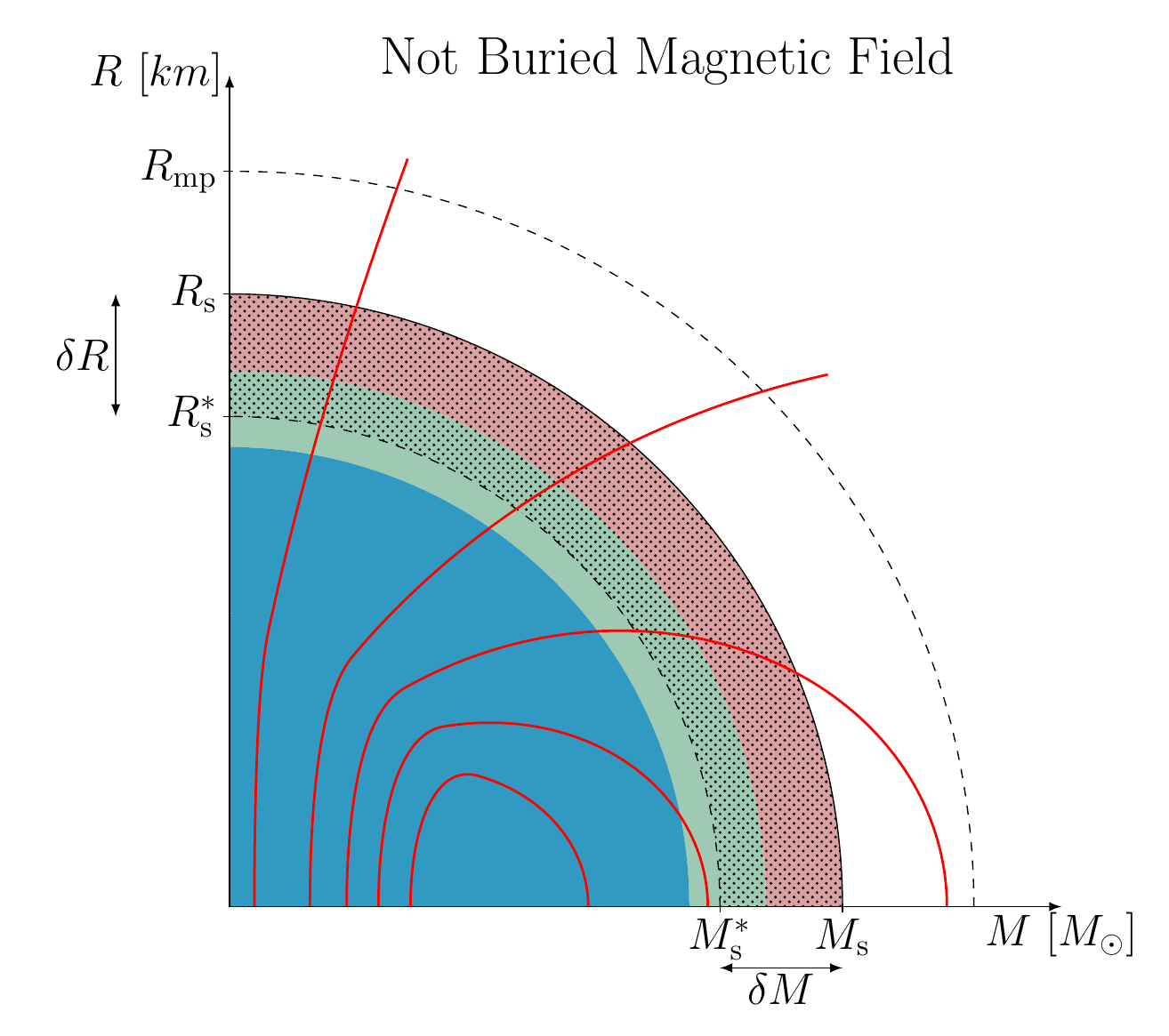}}
	\end{subfigure}\\

	\caption{Sketch of the representative stages of the accretion process. The upper panel shows the initial state of the process. The left column shows the expected evolutionary path for a low magnetic field ($B\lesssim 10^{13}$~G) while the right column correspond to a typical high magnetic field case (e.g.~$B\gtrsim 10^{13} $~G). A mass accretion rate of $10^{-5} M_{\odot}/$s is assumed. The scale ratio of the different regions is not preserved. See main text for details.}
	\label{fig:accretion_diagram}
\end{figure*}

\subsection{Setup}

Our goal is to study the conditions under which the magnetic field of a new-born NS can be buried by fallback material during a supernova. We have 
spanned a large range of values for both, the magnetic field  strength and the accretion rate, proceeding as follows.
We obtain the distance from the NS surface where the magnetosphere and the accreting fluid are in balance, i.e.~the radial point where the velocity of the contact discontinuity is zero. 
We reduce our 2D configuration to a 1D Riemann problem by  restricting the evaluation of the equilibrium point to the equatorial plane of the NS, due to the fact that the magnetic pressure is maximum at the equator. Therefore, if the magnetic field can be buried  in this latitude, it will be buried in all latitudes of the NS. 

The code developed by \cite{Romero:2005} requires as input the knowledge of the density, velocity, thermal pressure, and magnetic pressure at both left and right states of the initial discontinuity. In all cases we consider, the left state corresponds to the force-free magnetosphere while the right state is occupied by the accreting fluid. To obtain the magnetic pressure of the left state we find the solution of  the Grad-Shafranov equation (see Section~\ref{sec:forcefree}). This allows to locate the position of the magnetopause where the Riemann problem must be solved. Since the intertia of the fluid at the magnetosphere can be neglected in front of the magnetic pressure, the value of the density on the left state is set to yield an Alfv\'en velocity near to one , the thermal pressure is set to be at least six orders of magnitude lower than the magnetic pressure, and the velocity is set to zero. On the other hand, the values on the right state are fixed to the corresponding values of density, pressure and velocity of the stationary spherical accretion solution (see Section~\ref{sec:michel}) and the magnetic pressure is set to zero.
  
A sketch of the different stages of the accretion process is shown in Fig.~\ref{fig:accretion_diagram}. The plots depict the location of the NS (including its core and inner and outer crust), the magnetosphere, the magnetopause, and part of the region where material is falling back. Each region is shaded in a different colour for a simple identification. Note that the scale ratio of the different regions is not preserved in the figure.
The upper panel in Fig.~\ref{fig:accretion_diagram} shows the initial state of the process. The panels on the left column show the expected evolution for a low magnetic field case (e.g.~$B\lesssim 10^{13}$~G) while those on the right column correspond to a typical high magnetic field case (e.g.~$B\gtrsim 10^{13} $~G). In general, the value of $B$ separating between the two regimes depends on the 
accretion rate. For this figure we have chosen a value of the magnetic field that 
corresponds to a representative example of our results (see Section 6), for which $\dot{M}=10^{-5} M_{\odot}/$s.
At the beginning of the evolution, the reverse shock falls over the magnetosphere. The magnetic field lines are confined inside the magnetosphere, which is shown in white on the diagram.   
Depending on the position of the sonic point, which in turn depends on the values of the specific entropy and the accretion rate, the motion of the reverse shock may be either supersonic or subsonic.  
We limit the qualitative description of the evolution below to the case of a supersonic reverse shock as in the subsonic case no accretion shock forms, as shown in Section~\ref{sec:riemann}.

The middle two panels in both evolutionary tracks show only qualitative differences in the size of the resulting magnetosphere after its compression and in the amplitude of the instabilities that may arise in the magnetopause (see below). Therefore, our description can be used for either path keeping this quantitative differences in mind.
The evolution on the left column shows the case where the magnetic pressure is weak compared with the ram pressure of the fluid. In this case the magnetosphere shrinks significantly until the equilibrium point is reached ($R_{\rm mp}$; zero speed contact discontinuity) close to the NS surface at $R_{\rm s}\sim 10$~km. If the infall of the reverse shock is sxupersonic an accretion shock will appear simultaneously. The location of this accretion shock is shown on the horizontal axis of the four middle panels. As a result, the velocity of the reverse shock is reduced due to the presence of a region of subsonic accretion behind the accretion shock. Nevertheless, as through the accretion shock the momentum is conserved, the compression is not affected. The evolution on the right column, where the magnetic pressure is stronger, is qualitatively similar, only the accretion shock is located further away from the NS surface and the magnetosphere is not so deeply compressed.

As we will discuss below in more detail, the compression phase is unstable against the growth of Rayleigh-Taylor instabilities and the development of convection on the dynamic timescale. Therefore, the fluid and the magnetic field lines can mix, which provides a mechanism for the infalling fluid to actually reach the star.
As the fluid reaches the NS, the mass of the star grows from $M_{\rm s}^*$ to $M_{\rm s}$ and its radius increases from $R_{\rm s}^*$ to $R_{\rm s}$, encompassing the twisted magnetic field lines a short distance away. The mass accreted $\delta M$ forms part of the new crust of the NS, whose final radius will depend on the total mass accreted during the process.  
The bottom panels of the diagram depict a magnified view of the NS to better visualize the rearrangement the mass of the star and the magnetic field undergo.
If the radius $R_{\rm mp}$ of the equilibrium point is lower than the new radius $R_{\rm s}$, all the magnetic field lines will be frozen inside the NS new crust, as shown in the bottom-left plot of Fig.~\ref{fig:accretion_diagram} which corresponds to the end of the accretion process for a low magnetic field evolution. On the contrary, if the magnetic field is high, as considered on the evolutionary path on the right, the equilibrium point $R_{\rm mp}$ is far from the surface of the NS.  Although part of the infalling matter may still reach the star and form a new crust, the mechanism is not as efficient as in the low magnetic field case. This is depicted in the bottom-right panel of the figure. 

In our approach, that we discuss in more detail in the section on results, we compare the distance 
obtained by the Riemann solver for the location of $R_{\rm mp}$ (zero speed in the contact discontinuity) with the increment of the radius of the NS, $\delta R$, due to the pile up of the accreting matter. If the radial location of the equilibrium point $R_{\rm mp}$ is lower than $\delta R$ (as in the bottom-left panel of Fig.~\ref{fig:accretion_diagram}) we conclude that the magnetic field is completely buried into the NS crust. On the contrary, if $R_{\rm mp}>\delta R$, our approach does not allow us to draw any conclusion. In this case, multidimensional MHD numerical simulations must be performed to obtain the final state of the magnetic field.
\section{Results}
\label{Results}

We turn next to describe the main results of our study. In order to be as comprehensive as possible, we cover a large number of cases which are obtained from varying the physical parameters of the model, namely the composition and entropy of the accreting fluid, the mass of the NS, and the initial magnetic field distribution. For all possible combinations of these parameters the outcome of the accretion process depends both on the magnetic field strength and on the mass accretion rate. This dependence is presented in the following sections in a series of representative figures. A summary of all the combinations considered and the description of the model parameters can be found in Table~\ref{tab:models}.

\begin{table}
 \centering
  \caption{Models considered in this study.}
    \begin{tabular}{@{}cccccc@{}}  \hline
    \multicolumn{1}{|c|}
	{Model \#} & Composition & Entropy & NS Mass & MF distribution \\
	& &$k_{\rm{B}}  / \rm{nuc}$& $\rm{M}_\odot$&\\
        \hline
        Reference & He + NSE &10 & $1.4$& loop current \\
        1 & He + NSE &100 & $1.4$& loop current \\
        2 & He + NSE &1000 & $1.4$& loop current \\
        3 & He + NSE &5000 & $1.4$& loop current \\
        4 & He  &10 & $1.4$& loop current \\
        5 & He &100 & $1.4$& loop current \\
        6 & He &1000 & $1.4$& loop current \\
        7 & He & 5000 & $1.4$& loop current \\
	8 & C + NSE &10 & $1.4$& loop current \\
	9 & C + NSE &100 & $1.4$& loop current \\
	10 & C + NSE&1000 & $1.4$& loop current \\
	11 & C + NSE &5000 & $1.4$& loop current \\
	12 & He &10 & $1.4$& dipole \\
	13 & He &1000 & $1.4$& dipole \\
	14 & He &10& $1.2$& loop current \\
	15 & He &10& $1.6$& loop current \\
	16 & He &10& $1.8$& loop current \\
	17 & He &10& $2.0$& loop current \\

      \hline
\label{tab:models}
\end{tabular}
\end{table}

\subsection {Reference model}

We use as a reference model the one corresponding to an accreting fluid with $s=10~k_{\rm{B}}  / \rm{nuc}$, and composed essentially by Helium. The nuclear reactions to reach nuclear statistical equilibrium are also allowed in this model. The mass of the NS is $1.4\ M_{\odot}$ and the magnetic field is generated by a {\it loop current} in the NS. 
\begin{center}
	\begin{figure}
		{\includegraphics[width=80mm]{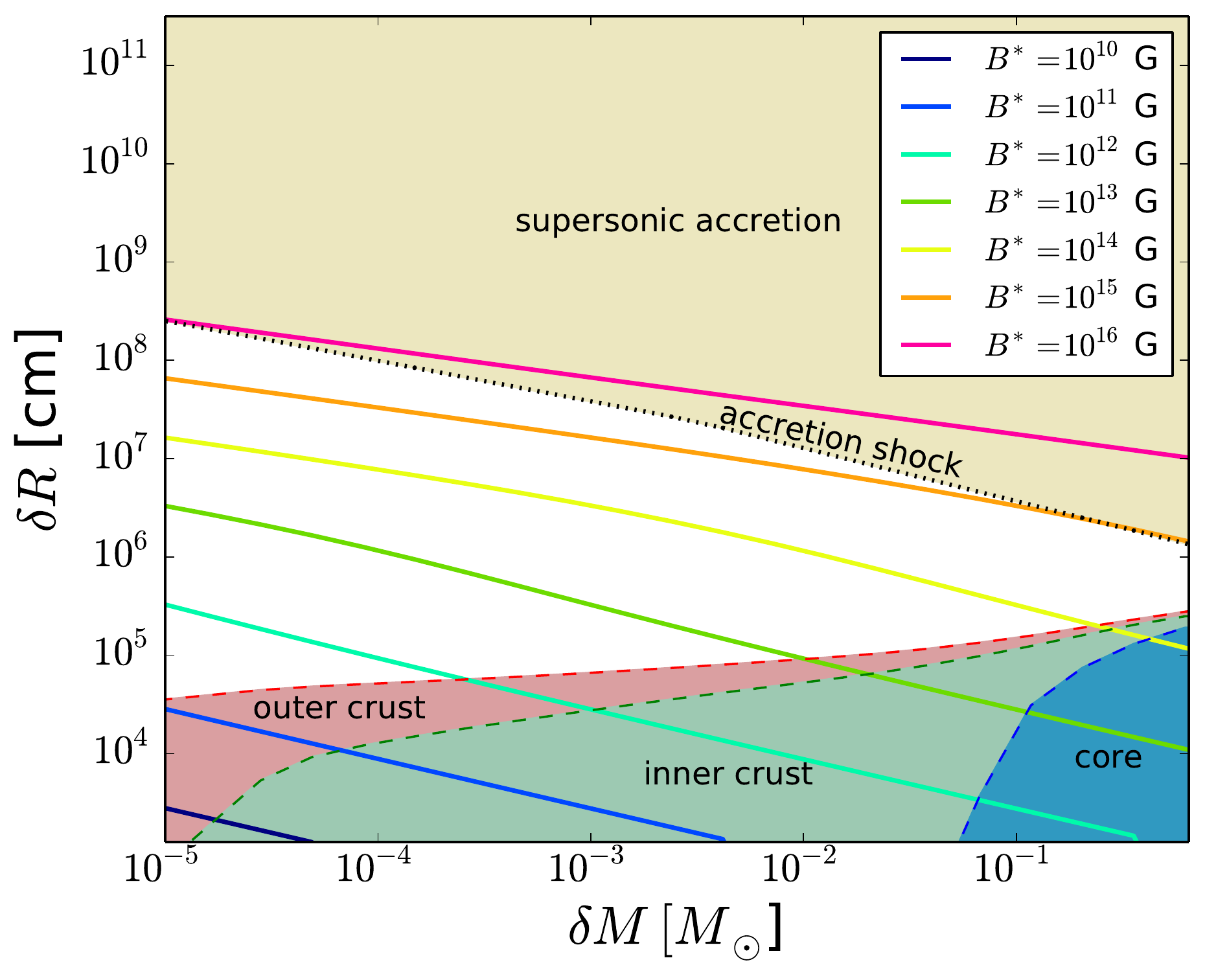}}
		\caption{Distance above the star of the equilibrium point $\delta R$ as a function of the total mass accreted $\delta M$ for each value of the magnetic field (solid lines) for the reference model. The yellow area indicates the region where the accretion flow is supersonic. The dotted line represents the limit of the accretion shock. The red area marks the outer crust of the NS after accretion, while the green and blue areas display the inner crust and the core respectively, as shown in Fig.~\ref{fig:accretion_diagram}.  }
		\label{fig:alternative_reference}
	\end{figure}
\end{center}
The results are shown in Fig.~\ref{fig:alternative_reference}. The solid lines in this figure represent the distance of the equilibrium point $\delta R$ (position of the magnetopause) above the NS surface as a function of the total accreted mass $\delta M$. The limit of the horizontal axis is given by the maximum mass that can be accreted without forming a black hole. Each line corresponds to a different value of the initial magnetic field, indicated in the legend of the figure.  The yellow area represents the region in which the accretion of the reverse shock is supersonic and the black dotted line shows the limit of the accretion shock. The dashed red line shows the radial location of the new surface of
the star due to the accretion of the infalling matter. The lines which cross the dashed red line have the equilibrium point inside the crust of the NS and, therefore, the corresponding magnetic fields
will be buried into the crust. However, for the lines that are in the white area, the equilibrium point is not close enough to the NS surface and the magnetic field can not be buried. Note that for initial values of the magnetic field $B\gtrsim 10^{15}$~G, the
magnetic field is never buried for all mass accretion rates considered. 
  
\begin{center}
\begin{figure}

	{\includegraphics[width=80mm]{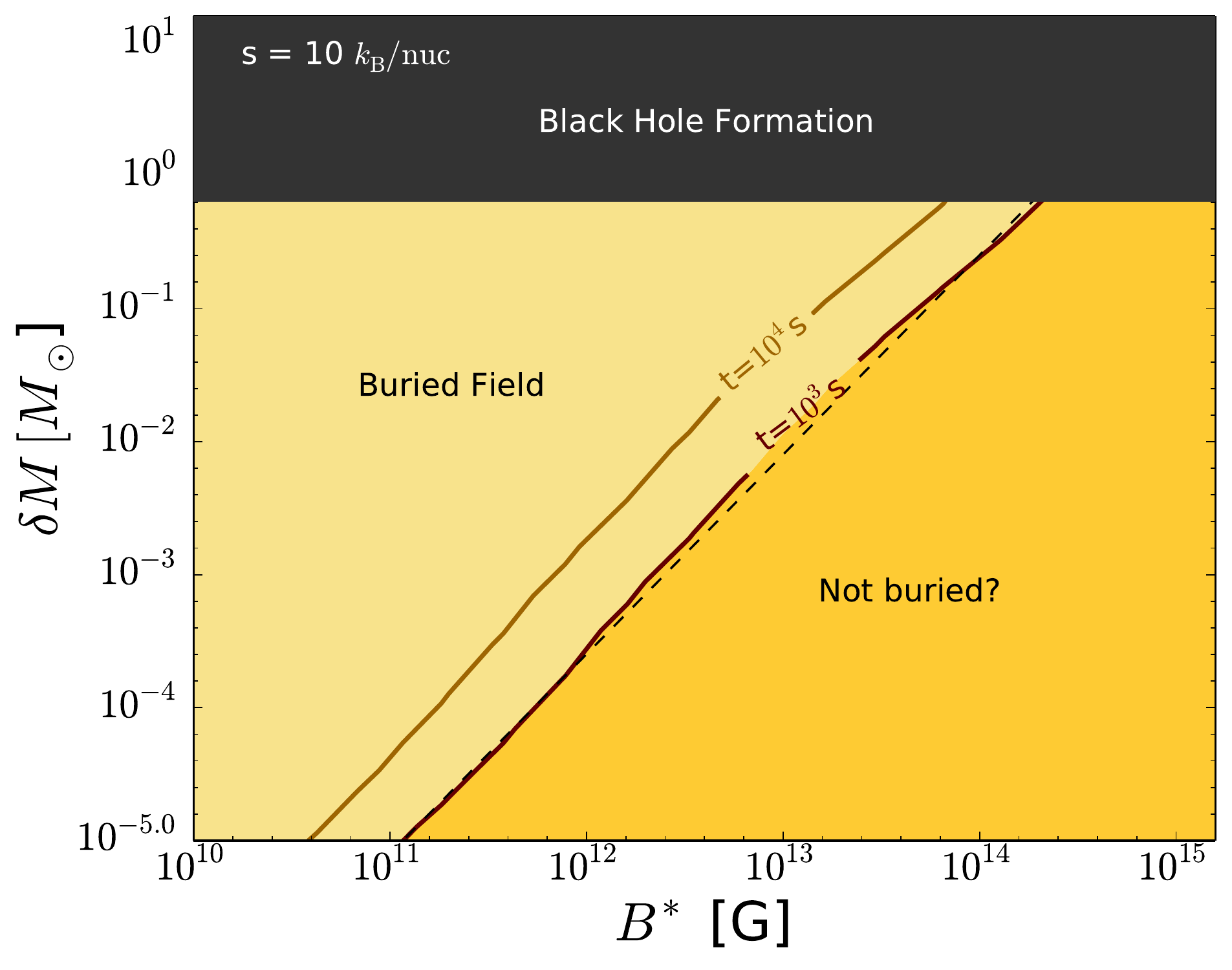}}

	\caption{Outcome of the accretion depending on the total accreted mass ($\delta M$) and the initial magnetic field ($B^*$) for the reference model.
          For the two accretion times considered, $t=10^{3}$~s (dark brown) and $t=10^4$~s (light brown), 
          the respective line splits the parameter space in a region where the magnetic field will be buried (left side) or not completely buried  (right side).
          Above certain $\delta M$ a black hole will be formed. The dashed line represents the fit shown in Eq.~(\ref{Eq:fit}).}
	\label{fig:reference}
\end{figure}
\end{center}

An alternative view of this result is shown in Fig.~\ref{fig:reference}. The goal of this representation is to provide a clearer representation of the dependence of the equilibrium point with the span of values of the magnetic field and the total mass accreted we are considering. The figure shows the isocontours where the equilibrium point is equal to the increment of the radius of the NS, i.e.~$R_{\rm mp}=\delta R$. The two lines plotted (dotted, $t=10^4$~s, and solid, $t=10^3$~s) correspond to the limits of the total accretion time, which relates the accretion rate $\dot{M}$ and the total mass increment $\delta M$. The black area indicates the values of the maximum mass of the NS beyond which it will form a black hole. The dark orange region represents the span of values of $\delta M$ and $B^*$ where we cannot assure that the magnetic field could be buried completely.
 The light orange area, on the other hand, represents the cases where the magnetic field is totally buried. 
The results show that for low values of the magnetic field ($B^*<10^{11} G$) the field can be buried even with the lowest accretion rates we have considered. As expected,
as the accreted mass increases it is possible to bury the magnetic field for larger initial field values, up to a certain maximum. Indeed, for $B^*> 2 \times 10^{14}$ G we cannot
find any accretion rate which can bury the magnetic field.

\subsection{Models with higher specific entropy}

\begin{figure}
	{\includegraphics[width=85mm]{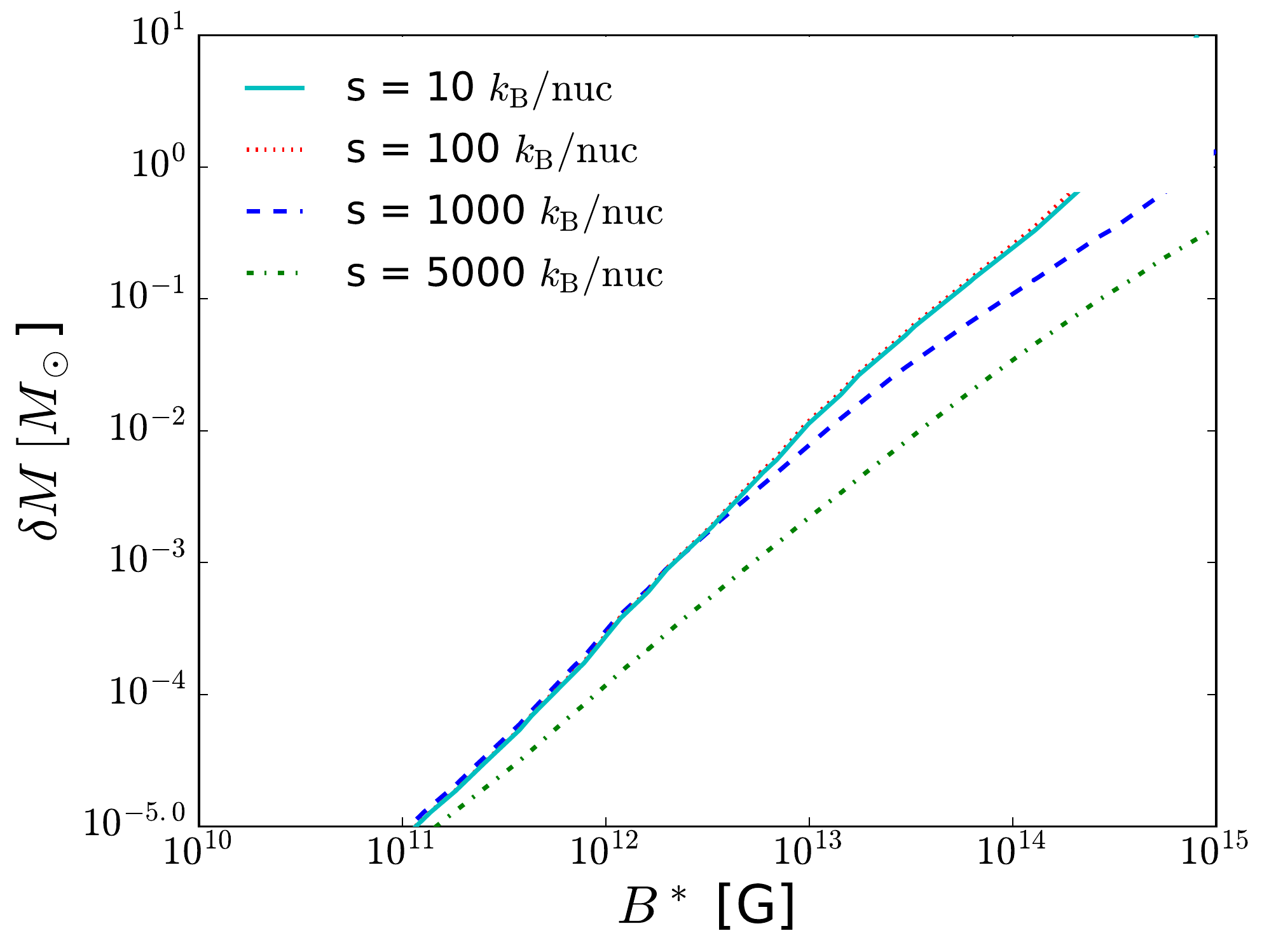}} \\
	\caption{Similar to Fig.~\ref{fig:reference} but for the models with different specific entropy for the accreting fluid (namely, models 1 to 3 and reference).
          All cases are shown for a total accretion time of $10^3$~s. Each line ends at the maximum mass of the corresponding model.}
	\label{fig:entropy_variation}
\end{figure}

We turn next to analyze the behavior of the magnetic field compression when the accreting fluid has higher specific entropy than in the reference model, keeping the same conditions for the composition, mass and magnetic field distribution (Models 1, 2 and 3 in Table~\ref {tab:models}). Fig.~\ref{fig:entropy_variation} shows the results for values of the specific entropy of $s=100~k_{\rm{B}}  / \rm{nuc}$, $1000~k_{\rm{B}}  / \rm{nuc}$ and $5000~k_{\rm{B}}  / \rm{nuc}$ compared with the reference model ($s=10~k_{\rm{B}}  / \rm{nuc}$). For the model with specific entropy $100~k_{\rm{B}}  / \rm{nuc}$, the results are very similar to the reference model as both lines almost perfectly overlap. For larger specific entropy the difference is more noticeable; for $s=1000~k_{\rm{B}}  / \rm{nuc}$
and  $5000~k_{\rm{B}}  / \rm{nuc}$, the burial/reemergence boundary of the parameter space is shifted toward larger magnetic fields, i.e. higher entropy material compress the magnetosphere more easily and it is possible to bury larger magnetic fields. This behavior can be understood if one considers that the equilibrium
point is a balance between the total pressure of the infall material, $p_{\rm tot} = p + p_{\rm ram} \approx p + \rho v^2$, and the magnetic pressure of the
magnetosphere. For low specific entropy, the total pressure is dominated by the ram pressure and changes in $s$ do not produce significant changes in the
equilibrium point. Above a certain threshold, the thermal pressure $p$ dominates the total pressure and increasing $s$ induces a larger compression of the
magnetosphere, shifting the equilibrium point downwards. For realistic values of the specific entropy in supernovae, $s\sim 10-100~k_{\rm{B}}  / \rm{nuc}$ \citep{Scheck:2006,Kifonidis:2003,Kifonidis:2006}, we expect the ram pressure to be dominant and hence the influence of $s$ to be minimal. Even for an unrealistically
large value of the specific entropy, $5000~k_{\rm{B}}  / \rm{nuc}$, the maximum magnetic field that can be buried increases one order of magnitude at most,
and only for the largest mass accretion rates considered.

\subsection{Models with different NS mass}

\begin{figure}
	{\includegraphics[width=85mm]{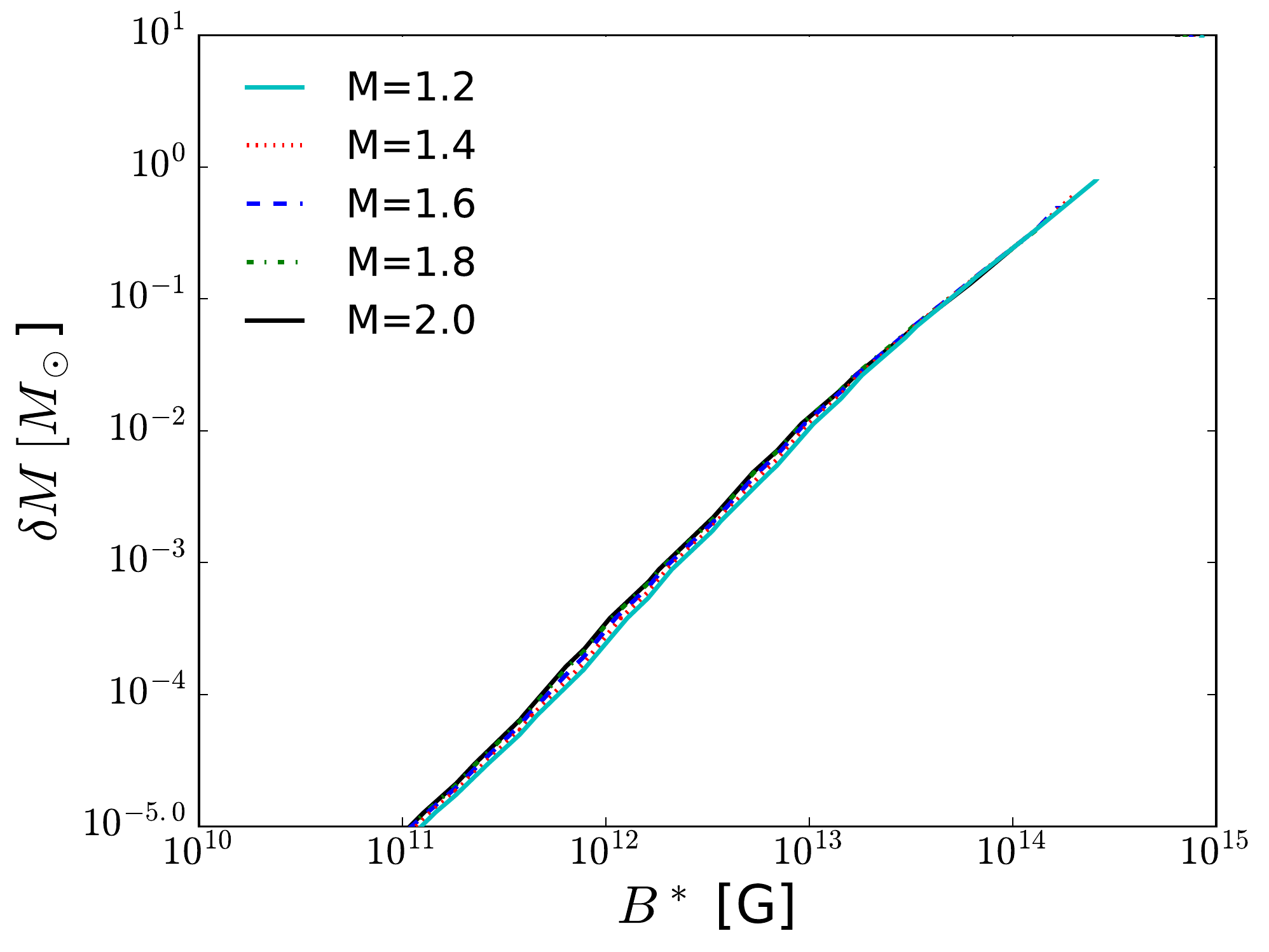}}
	\caption{Similar to Fig.~\ref{fig:reference} for the models with different NS mass: $1.2~{M}_\odot$, $1.6~{M}_\odot$ ,
          $1.8~{M}_\odot$, $2.0~{M}_\odot$  (models 14, 15, 16 and 17 and reference).
          All cases are shown for a total accretion time of $10^3$~s. Each line ends at the maximum mass of the corresponding model.}
	\label{fig:mass_variation}
\end{figure}

We consider next the effect of the neutron star mass, within astrophysically relevant limits.
According to observations \citep[see][and references therein]{Lattimer:2012} the lower limit for the NS mass is around $1.2~M_\odot$. The maximum achievable mass of a NS is strongly dependent on the equation of state \citep{Lattimer:2005}. Nowadays, there are a few observations that support the existence of pulsars and NS with masses greater than $1.5~M_\odot$, in particular an observation of a $\sim 2~M_\odot$ NS \citep{Demorest:2010,Antoniadis:2013}. For this reason, we explore the results for several values of the neutron star mass between $1.2~M_\odot$ and $2~M_\odot$. The results are shown in Fig. \ref{fig:mass_variation}, where each line corresponds to a model with different NS mass as indicated in the legend.
The results for all masses are very similar. In general we observe that for more massive NS, a higher accreted mass is needed to bury the
  magnetic field. Our interpretation is that higher mass NS have lower radii and hence we have to compress more the magnetosphere to successfully
  bury it into the crust. Therefore, a higher accreted mass is needed to bury the field for NS with larger mass (smaller radius). Since the radius
  difference between a $1.2$ and a $2$~$M_\odot$ NS is small, the impact of the NS mass on the burial is minimal.
The maximum value of the magnetic field which can be buried is $\sim 2\times10^{14}$ G in all cases. For smaller NS masses slightly larger values of the magnetic field can be buried due to the ability to support a larger accreted mass. We conclude that the burial of the magnetic field is not crucially sensitive to the NS mass. 
%
\subsection{Models with different EoS}

\begin{figure}
	{\includegraphics[width=85mm]{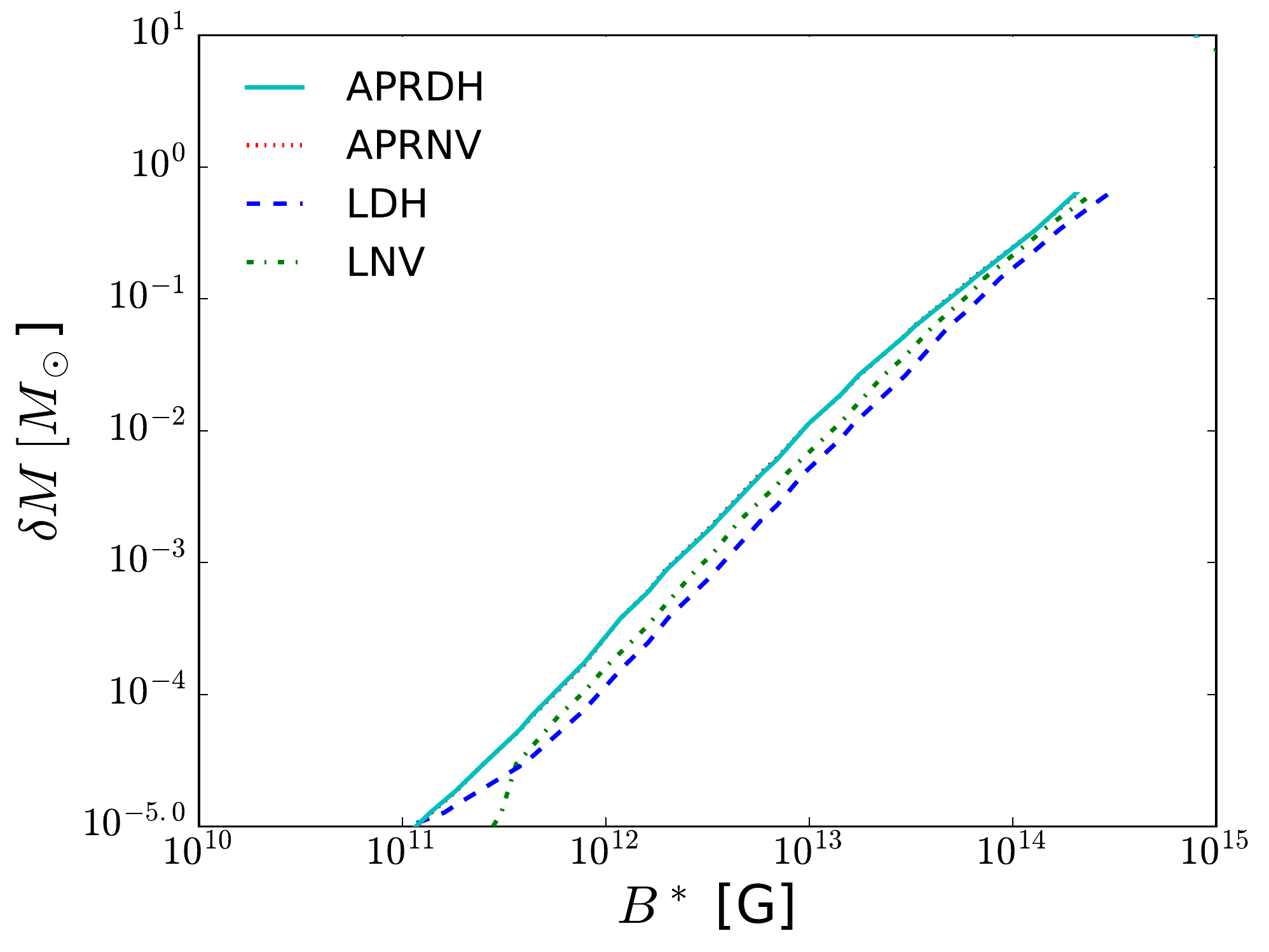}}
	\caption{Similar to Fig.~\ref{fig:reference} for the models with different EoS,  a NS of mass $1.4\rm{M}_\odot$ and
          specific entropy of the accreting fluid $10k_{\rm{B}}  / \rm{nuc}$ 
          All cases are shown for a total accretion time of $10^3$~s. Each line ends at the maximum mass of the corresponding model. }
	\label{fig:eos_variation}
\end{figure}

Fig.~\ref{fig:eos_variation} shows the comparison of the results for the reference model when using the four different equations of state described in 
section~\ref{sec:non-magnetized accretion}. For $M=1.4~M_\odot$, the coordinate radius of these NS models is $12.25$~km for APRDH, $12.11$~km for APRNV,
$15.77$~km for LDH and $15.37$~km for LNV.
Since the maximum mass is sensitive to the EoS, each line ends at different points in the $\delta M$ vs $B^*$ plot. The use of APRDH or APRNV EoSs leads to almost indistinguishable results (the two lines
lay on top of each other). This is expected since the radius of this two models differs only by about $1\%$, because the EoS are very similar and only
differ at low densities (at the crust). The LDH and LNV EoSs allow the burial of a larger magnetic field for a given accreted mass, in comparison with
APRDH and APRNV. The maximum magnetic field that can be buried in the LDH and LNV models is $\sim 6\times10^{14}$ G and $\sim 5\times10^{14}$ G respectively,
which is about a factor $2$ larger than for the APRDH EoS. In general, for a $M=1.4~M_\odot$, EoS resulting in a larger NS radius allow to bury larger magnetic fields for
a given $\delta M$. Given that the results of this work are meant to be an order-of-magnitude estimate of the location in the parameter space of the limit between
burial and reemergence, a difference of a factor $2$ due to the EoS, does not change the main conclusions of this work. For practical purposes the APRDH EoS
can be taken as a good estimator for this limit.

\subsection{Remaining models}

We do not observe any significant differences with respect to the reference model in the results for the models with different initial composition of the reverse shock (models 8 to 11) or the ones using the NSE calculations (models 4 to 7). As a result we do not present additional figures for these models since the limiting lines overlap with those of the reference model. 
The observed lack of dependence is due to the fact that the EoS only depends on the electron fraction, $Y_e$. This value is obtained from the ratio between the mean atomic mass number ($\bar{A}$) and the mean atomic number ($\bar{Z}$). For both cases of pure Helium and pure Carbon, this ratio is equal to $Y_e=0.5$ and, consequently, the values of pressure and density for the accreting fluid are almost identical, producing differences in the results below the numerical error of our method~\footnote{The numerical error is dominated by the calculation of the equilibrium point, which is computed with a relative accuracy of $10^{-4}$.}. In the case of the NSE calculation, the reason is similar. For low entropies ($s=10, 100\,k_{\rm{B}}  / \rm{nuc}$) the temperature is not sufficiently high to start the nuclear reactions and the composition remains constant throughout the accretion phase. For higher entropies, although the value of the electron fraction may differ from $0.5$ during the accretion process, the differences produced in the thermodynamical variables lead to changes in the results of the Riemann problem still below the numerical error of the method.

Regarding the initial distribution of the magnetic field, we do not observe either any significant difference in the results in the two cases that we have considered, {\it loop current} and {\it dipole}. Given that we are comparing models with the same effective magnetic field, $B^*$, and thus the same magnetic dipolar moment, the magnetic field is virtually identical at long radial distances and the only differences appear close to the NS surface. In practice the  magnetic field structure only changes the details of the burial
in the cases in which the equilibrium point is close to the burial depth (the limiting line plotted in the Figs.~\ref{fig:reference} to \ref{fig:eos_variation}),
but it does not change the location of the limit itself in a sensitive way. As a conclusion, we can say that the dominant ingredient affecting the burial of the magnetic
field is the presence of a dipolar component of the magnetic field but, for order-of-magnitude estimations, a multipolar structure of the field is mostly irrelevant.

\section{Summary and discussion}
\label{Summary}

We have studied the process of submergence of magnetic field in a newly born neutron star during a hypercritical accretion stage in coincidence with core collapse supernovae explosions. This is one of the possible scenarios proposed to explain the apparently low external dipolar field of CCOs. Our approach is based on 1D solutions of the relativistic Riemann problem, which provide the location of the spherical boundary (magnetopause) matching an external non-magnetised accretion solution with an internal magnetic field potential solution. For a given accretion rate and magnetic field strength, the magnetopause keeps moving inwards if the total (matter plus ram) pressure of the accreting fluid, exceeds the magnetic pressure below the magnetopause. Exploring a wide range of accretion rates and field strengths, we have found the conditions for the magnetopause to reach
the equilibrium point below the NS surface, which implies the burial of the magnetic field.
Our study has considered several models with different specific entropy, composition, and neutron star masses. Assuming an accretion time of 1000s, our findings can be summarised by a general condition, rather independent on the model details, relating the required total accreted mass to bury the magnetic field with the field strength. An approximate fit is (see dashed line in Fig.~\ref{fig:reference})
\begin{equation}
\label{Eq:fit}
\frac{\delta M}{M_\odot} \approx \left( \frac{B}{2.5\times10^{14}} \right)^{2/3}.
\end{equation}

The most important caveat in our approach is that we are restricted to a simplistic 1D spherical geometry, which does not allow us to consistently account for the effect of different MHD instabilities that can modify the results. We also note that our scenario is quite different from the extensively studied case of X-ray binaries, in which the NS accretes 
matter from a companion but at much lower rates (sub-Eddington) and matter is mostly transparent to radiation during accretion. In that case, matter cools down through X-ray emission during the accretion process. \cite{Davidson:1973} and \cite{Lamb:1973} already noticed this fact and predicted that the accretion will most likely be channeled through the magnetic poles, in analogy to the Earth's magnetosphere. In the context of X-ray binaries, 
\cite{Arons:1976} and \cite{Michel:1977} were able to compute equilibrium solutions with a deformed magnetosphere and a cusp like accretion region at the magnetic poles. However, as the same authors pointed out, these systems are unstable to the interchange instability \citep{Kruskal:1954}, a Railegh-Taylor-like instabilitiy in which magnetic field flux tubes from the magnetosphere can raise, allowing the fluid to sink. This might allow for the formation of bubbles of material that fall through the magnetosphere down to the NS surface. In the case of a fluid deposited on top of a highly magnetized region, modes with any possible wavelength will be unstable \citep{Kruskal:1954}, however, in practice these instabilities are limited to the size of the magnetosphere ($\sim R_{\rm mp}$) in the angular direction. As the bubbles of accreted material sink, magnetic flux tubes raise, as long as their magnetic pressure equilibrates the ram pressure of the unmagnetized accreting fluid \citep{Arons:1976}. Therefore, in a natural way, the equilibrium radius computed in Section~\ref{sec:riemann} roughly determines the highest value at which the magnetic field can raise. 

This accretion mechanism through instabilities has been shown to work in the case of X-ray binaries in global 3D numerical simulations \citep[e.g.][]{Kulkarni:2008,Romanova:2008}.
In the case of the hypercritical accretion present in the supernova fallback, Rayleigh-Taylor instabilities have been studied by \cite{Payne:2004, Payne:2007,Bernal:2010, Bernal:2013,Mukherjee:2013a, Mukherjee:2013b}. The simulations of \cite{Bernal:2013} also show that the height of the unstable magnetic field over the NS surface decreases with increasing accretion rate, for fixed NS magnetic field strength, as expected. Using the method described in Section~\ref{sec:riemann} we have estimated the equilibrium height over the NS surface for the 4 models presented in Fig.~9 of \cite{Bernal:2013}, for their lower accretion rates ($\dot{M}\le 10^{-6}\ \rm{M}_\odot/s$). Our results predict correctly the order of magnitude of the extent of the unstable magnetic field
over the NS surface. 
Therefore, our simple 1D model for the equilibrium radius serves as a good estimator of the radius confining
the magnetic field during the accretion process, although details about the magnetic field structure cannot be predicted.
Another important difference with the binary scenario is the duration of the accretion process. In X-ray binaries, a low accretion rate is maintained over
very long times, so that instabilities have always time to grow. In our case, hypercritical accretion can last only hundreds or thousands of seconds, and depending on the particular values of density and magnetic field, this may be too short for some instabilities to fully develop. This issue is out of the scope of this paper and deserves a more detailed study. 

Our main conclusion is that a typical magnetic field of a few times $10^{12}$ G can in principle be buried by accreting only $10^{-3}-10^{-2} M_\odot$, 
a relatively modest amount of mass. This estimate has interesting implications: since it is likely that most neutron stars can undergo 
such an accretion process, and the field would only reemerge after a few thousand years \citep {Geppert:1999, Vigano:2012}, the CCO scenario is 
actually not peculiar at all and we expect that most very young NSs show actually an anomalously low value of the magnetic field. 
On the contrary, magnetar-like field strengths are much harder to screen and the required accreted mass is very large, in some cases so large that the neutron star would collapse to a black hole. We also stress that the concept of {\it burial} of the magnetic field refers only to the large scale dipolar component, responsible for the magnetospheric torque spinning down the star. Small scale structures produced by instabilities can exist in the vicinity of the star surface, and this locally strong field is likely to have a visible imprint in the star thermal spectrum, as in Kes 79 \citep{Shabaltas:2012}, without modifying the spin-down torque.
However, the high field burial scenario should not be
very common because both, high field NSs are only a fraction to the entire population, and only a part of them would undergo the fallback episode with the right amount of matter.
This is consistent with the recent results of \cite{Bogdanov:2014} who searched for the hidden population of evolved CCOs among a sample of normal pulsars with old characteristic ages but close to a supernova remnant. None of the eight sources studied
was found to have a luminosity higher than $10^{33}$ erg/s, which would have been an evidence of a hidden strong field. They all show X-ray luminosities in the 0.3-3 keV band of the order of $10^{31}$ erg/s  (or similar upper limits),
consistent with the properties of other low field neutron stars with $B\approx 10^{12}$ G. Thus, these sample of sources are not likely to be linked to the family of descendants of Kes 79-like objects, but there is no contradiction with these being pulsars with reemerged normal fields.
Finally, we note that the slow reemergence process on timescales of kyrs mimics the increase of the magnetic field strength, and it is therefore consistent with a value of the braking index smaller than 3 \citep{Espinoza:2011}, which should be common for all young pulsars in this scenario.


\section*{Acknowledgments}
It is a pleasure to thank J.M. Mart\'i for many fruitful discussions.
This work has been supported by the Spanish MINECO grants AYA2013-40979-P and  AYA2013-42184-P and 
by the Generalitat Valenciana (PROMETEOII-2014-069).

\appendix

\section{Riemann Problem}
\label{riemann_problem}
For an ideal magneto-fluid, the energy-momentum tensor $T^{\mu\nu}$ and Maxwell dual tensor $F^{*\mu \nu}$ are
\begin{eqnarray}
T^{\mu \nu}&=&\rho\hat{h}u^{\mu}u^{\nu}+g^{\mu \nu}\hat{p}-b^\mu b^\nu\,,
\\
F^{*\mu \nu} &=& u^\mu b^\nu-u^\nu b^\mu\,,
\end{eqnarray}
where $\hat{h}=1+\varepsilon+p/\rho+b^2/\rho$ is the specific enthalpy including the contribution of the magnetic field and $\hat{p}=p+b^2/2$ is the total pressure. Moreover, $b^{\mu}$ stands for the magnetic field measured by a comoving observer (see~\cite{Anton:2006} for details and its relation with the magnetic field $B^{\mu}$ measured by a generic observer). The conservation of these two quantities, jointly with the conservation of the density current, equation~(\ref{eq:density_current}), lead to the equations of ideal relativistic MHD. 
\begin{eqnarray}
\nabla_{\mu} J^\mu=0\,,
\\
\nabla_{\mu}T^{\mu \nu}=0\,,
\\
\nabla_{\mu}F^{*\mu\nu}=0\,.
\end{eqnarray}
In the particular configuration of our Riemann problem, $u^\mu=W(1,v^x,0,v^z)$, $b^\mu=(0,0,b,0)$, so the term $\nabla_{\mu}b^\mu b^\nu$ in the conservation of the stress-energy tensor, vanish. Therefore, the conservation equations reduce to the purely hydrodynamical case. 

The Riemann problem in this particular configuration is described in terms of three characteristics, one entropy wave and two fast magnetosonic waves.
The initial problem with two states $L$ (left) and $R$ (right) breaks up into four states,
\begin{eqnarray}
L\mathcal{W}_\leftarrow L_*\mathcal{C} R_* \mathcal{W}_\to R,
\end{eqnarray}
where $\mathcal{W}$ indicate a fast magnetosonic shock wave or a rarefaction wave and $\mathcal{C}$ indicates the contact discontinuity. Solving the Riemann problem entails finding the intermediate states ($L_*,R_*$) and the position of the waves, which are determined by the pressure $\hat{p}_*$ and the flow velocity $v_*^x$. If $\hat{p}\leq\hat{p}_*$ the wave is a rarefaction wave (a self-similar continuous solution), otherwise the solution is a shock wave. In our case \citep[see][]{Romero:2005} the ordinary differential equation that allows to obtain the solution for a rarefaction wave is given by
\begin{eqnarray}
\label{eq:rarefaction}
\frac{{d}v^x}{1-(v^x)^2}=\pm \frac{(1+b^2/(\rho h c_s))\sqrt{\hat{h}+\hat{\mathcal{A}}^2(1-w^2)}}{\hat{h}^2+\hat{\mathcal{A}}^2}\frac{{d}p}{\rho w},
\end{eqnarray}
where $\hat{\mathcal{A}}=\hat{h}Wv^z$, $w=c_s^2+v_A^2-c_s^2v_A^2$, $v_A=b^2/\rho \hat{h}$ is the Alfv\'en velocity and $c_s=\sqrt{\frac{1}{h}\left.\frac{\partial p}{\partial \rho}\right|_s}$ is the sound speed. The integration of equation~\ref{eq:rarefaction} allows to conect the states ahead (a) and behind (b) the rarefaction wave. Rarefaction waves conserve entropy, hence, all the thermodynamical variables and the differential of $p$ must be calculated at the same entropy of the initial state. From this equation, the normal velocity behind the rarefaction can be obtained directly,
\begin{eqnarray}
v_b^x=\tanh \hat{\mathcal{C}},
\end{eqnarray}
with
\begin{eqnarray}
\hat{\mathcal{C}}&=& \frac{1}{2}\log\left(\frac{1+v_a^x}{1-v_a^x}\right)
\nonumber \\ 
&\pm&
\int_{\hat{p}_a}^{\hat{p}_b}\frac{(1+b^2/(\rho h c_s))\sqrt{\hat{h}+\hat{\mathcal{A}}^2(1-w^2)}}{\hat{h}^2+\hat{\mathcal{A}}^2}\frac{{d}p}{\rho w}.
\end{eqnarray}
In the same way, the velocity inside the rarefaction can be obtained replacing the thermal pressure  $p$ by the total pressure $\hat{p}$.

On the other hand, shock waves should fulfill the so-called Rankine-Hugoniot conditions \citep{Lichnerowicz:1967,Anile:1989}
\begin{eqnarray}
[\rho u^{\mu}] \,n_{\mu} &=& 0\,,
\\
{[ T^{\mu\nu}]} \,n_{\nu} &=& 0\,,
\\
{[F^{*\mu \nu}]} \,n_{\nu} &=& 0\,,
\end{eqnarray}
where $n_\mu$ is the unit normal to a given surface and $[H]\equiv H_a-H_b$ being $H_a$ and $H_b$ the boundary values. The normal flow speed in the post-shock state, $v_b^x$, can be extracted from the Rankine-Hugoniot equations \citep[see][for a detailed discussion]{Romero:2005},
\begin{eqnarray}
v_b^x&=&\left(\hat{h}_aW_av_a^x+\frac{W_s(\hat{p}_b-\hat{p}_a)}{j} \right)
\\
&\times& 
\left(\hat{h}_aW_a +(\hat{p}_b-\hat{p}_a)\left(\frac{W_sv_a^x}{j}+\frac{1}{\rho_a W_a}\right)\right)^{-1}\,,
\end{eqnarray}
where $W_s=\frac{1}{\sqrt{1-V_s^2}}$ is the Lorentz factor of the shock, 
\begin{eqnarray}
V_s^{\pm}=\frac{\rho_a^2W_a^2v_a^x\pm|j|\sqrt{j^2+\rho_a^2W_a^2(1-{v_a^x})^2}}{\rho_a^2W_a^2+j^2}\,,
\end{eqnarray}
is the shock speed, and 
\begin{eqnarray}
j\equiv W_s\rho_aW_a(V_s-v_a^x)=W_s\rho_bW_b(V_s-v_b^x)\,,
\end{eqnarray}
is an invariant derived directly from the Rankine-Hugoniot jump conditions. 
These expressions, together with the Lichnerowicz adiabat,
\begin{eqnarray}
[\hat{h}^2]=\left(\frac{\hat{h}_b}{\rho_b}+\frac{\hat{h}_a}{\rho_a}\right),
\end{eqnarray}
allows us to calculate the shock wave solution.

\end{document}